\documentclass[iop]{emulateapj}

\usepackage{mathptmx} 
\usepackage{lineno} 
\usepackage{amsmath,amssymb,amstext}
\usepackage{multirow}
\usepackage[breaklinks,colorlinks,citecolor=blue,linkcolor=blue,urlcolor=blue]{hyperref}
\usepackage{siunitx} 
\DeclareSIUnit\solarluminosity{\ensuremath{L_\sun}}
\DeclareSIUnit\solarmass{\ensuremath{M_\sun}}

\definecolor{ApJPink}{RGB}{255,0,255} 
\newcommand{\mydoi}[2]{\href{#1}{\color{ApJPink}{#2}}}
\newcommand{\species}[2]{#1\,\textsc{#2}}
\newcommand{\myemail}{voge@physik.uni-bonn.de}

\usepackage{etoolbox}
\makeatletter
\patchcmd{\NAT@citex}
  {\@citea\NAT@hyper@{%
     \NAT@nmfmt{\NAT@nm}%
     \hyper@natlinkbreak{\NAT@aysep\NAT@spacechar}{\@citeb\@extra@b@citeb}%
     \NAT@date}}
  {\@citea\NAT@nmfmt{\NAT@nm}%
   \NAT@aysep\NAT@spacechar\NAT@hyper@{\NAT@date}}{}{}
\patchcmd{\NAT@citex}
  {\@citea\NAT@hyper@{%
     \NAT@nmfmt{\NAT@nm}%
     \hyper@natlinkbreak{\NAT@spacechar\NAT@@open\if*#1*\else#1\NAT@spacechar\fi}%
       {\@citeb\@extra@b@citeb}%
     \NAT@date}}
  {\@citea\NAT@nmfmt{\NAT@nm}%
   \NAT@spacechar\NAT@@open\if*#1*\else#1\NAT@spacechar\fi\NAT@hyper@{\NAT@date}}
  {}{}
\makeatother

\shorttitle{Supernova Detection in IceCube follow-up}
\shortauthors{M.~G.~Aartsen et al.}

\begin{document}

\journalinfo{Preprint accepted for publication in The Astrophysical Journal}
\submitted{}

\title{Detection of a Type IIn Supernova in Optical Follow-up Observations\\of IceCube Neutrino Events}

\author{M.~G.~Aartsen\altaffilmark{1},
K.~Abraham\altaffilmark{2},
M.~Ackermann\altaffilmark{3},
J.~Adams\altaffilmark{4},
J.~A.~Aguilar\altaffilmark{5},
M.~Ahlers\altaffilmark{6},
M.~Ahrens\altaffilmark{7},
D.~Altmann\altaffilmark{8},
T.~Anderson\altaffilmark{9},
M.~Archinger\altaffilmark{10}
% ...
\\
(IceCube Collaboration)
}
\affil{$^{1}$ School of Chemistry \& Physics, University of Adelaide, Adelaide SA, 5005 Australia\\
$^{2}$ Technische Universit\"at M\"unchen, D-85748 Garching, Germany\\
$^{3}$ DESY, D-15735 Zeuthen, Germany\\
$^{4}$ Dept.~of Physics and Astronomy, University of Canterbury, Private Bag 4800, Christchurch, New Zealand\\
$^{5}$ Universit\'e Libre de Bruxelles, Science Faculty CP230, B-1050 Brussels, Belgium\\
$^{6}$ Dept.~of Physics and Wisconsin IceCube Particle Astrophysics Center, University of Wisconsin, Madison, WI 53706, USA\\
$^{7}$ Oskar Klein Centre and Dept.~of Physics, Stockholm University, SE-10691 Stockholm, Sweden\\
$^{8}$ Erlangen Centre for Astroparticle Physics, Friedrich-Alexander-Universit\"at Erlangen-N\"urnberg, D-91058 Erlangen, Germany\\
$^{9}$ Dept.~of Physics, Pennsylvania State University, University Park, PA 16802, USA\\
$^{10}$ Institute of Physics, University of Mainz, Staudinger Weg 7, D-55099 Mainz, Germany\\
$^{11}$ III\@. Physikalisches Institut, RWTH Aachen University, D-52056 Aachen, Germany\\
$^{12}$ Physics Department, South Dakota School of Mines and Technology, Rapid City, SD 57701, USA\\
$^{13}$ Dept.~of Physics and Astronomy, University of California, Irvine, CA 92697, USA\\
$^{14}$ Dept.~of Physics, University of California, Berkeley, CA 94720, USA\\
$^{15}$ Dept.~of Physics and Center for Cosmology and Astro-Particle Physics, Ohio State University, Columbus, OH 43210, USA\\
$^{16}$ Dept.~of Astronomy, Ohio State University, Columbus, OH 43210, USA\\
$^{17}$ Fakult\"at f\"ur Physik \& Astronomie, Ruhr-Universit\"at Bochum, D-44780 Bochum, Germany\\
$^{18}$ Dept.~of Physics, University of Wuppertal, D-42119 Wuppertal, Germany\\
$^{19}$ Dept.~of Physics, University of Maryland, College Park, MD 20742, USA\\
$^{20}$ Dept.~of Physics and Astronomy, University of Kansas, Lawrence, KS 66045, USA\\
$^{21}$ Lawrence Berkeley National Laboratory, Berkeley, CA 94720, USA\\
$^{22}$ Dept.~of Physics and Astronomy, Uppsala University, Box 516, S-75120 Uppsala, Sweden\\
$^{23}$ Dept.~of Physics, TU Dortmund University, D-44221 Dortmund, Germany\\
$^{24}$ Dept.~of Physics, Sungkyunkwan University, Suwon 440-746, Korea\\
$^{25}$ Vrije Universiteit Brussel, Dienst ELEM, B-1050 Brussels, Belgium\\
$^{26}$ Dept.~of Physics, University of Alberta, Edmonton, Alberta, Canada T6G 2E1\\
$^{27}$ School of Physics and Center for Relativistic Astrophysics, Georgia Institute of Technology, Atlanta, GA 30332, USA\\
$^{28}$ D\'epartement de physique nucl\'eaire et corpusculaire, Universit\'e de Gen\`eve, CH-1211 Gen\`eve, Switzerland\\
$^{29}$ Dept.~of Physics, University of Toronto, Toronto, Ontario, Canada, M5S 1A7\\
$^{30}$ Dept.~of Astronomy and Astrophysics, Pennsylvania State University, University Park, PA 16802, USA\\
$^{31}$ Dept.~of Physics and Astronomy, Michigan State University, East Lansing, MI 48824, USA\\
$^{32}$ Bartol Research Institute and Dept.~of Physics and Astronomy, University of Delaware, Newark, DE 19716, USA\\
$^{33}$ Dept.~of Physics and Astronomy, University of Gent, B-9000 Gent, Belgium\\
$^{34}$ Institut f\"ur Physik, Humboldt-Universit\"at zu Berlin, D-12489 Berlin, Germany\\
$^{35}$ Dept.~of Physics, Southern University, Baton Rouge, LA 70813, USA\\
$^{36}$ Dept.~of Physics, Chiba University, Chiba 263-8522, Japan\\
$^{37}$ Dept.~of Astronomy, University of Wisconsin, Madison, WI 53706, USA\\
$^{38}$ Physikalisches Institut, Universit\"at Bonn, Nussallee 12, D-53115 Bonn, Germany\\
$^{39}$ CTSPS, Clark-Atlanta University, Atlanta, GA 30314, USA\\
$^{40}$ Department of Physics, Yale University, New Haven, CT 06520, USA\\
$^{41}$ Dept.~of Physics and Astronomy, Stony Brook University, Stony Brook, NY 11794-3800, USA\\
$^{42}$ Universit\'e de Mons, 7000 Mons, Belgium\\
$^{43}$ Niels Bohr Institute, University of Copenhagen, DK-2100 Copenhagen, Denmark\\
$^{44}$ Dept.~of Physics, Drexel University, 3141 Chestnut Street, Philadelphia, PA 19104, USA\\
$^{45}$ Dept.~of Physics, University of Wisconsin, River Falls, WI 54022, USA\\
$^{46}$ Dept.~of Physics and Astronomy, University of Alabama, Tuscaloosa, AL 35487, USA\\
$^{47}$ Dept.~of Physics and Astronomy, University of Alaska Anchorage, 3211 Providence Dr., Anchorage, AK 99508, USA\\
$^{48}$ Dept.~of Physics, University of Oxford, 1 Keble Road, Oxford OX1 3NP, UK\\
$^{49}$ Earthquake Research Institute, University of Tokyo, Bunkyo, Tokyo 113-0032, Japan\\
$^{50}$ NASA Goddard Space Flight Center, Greenbelt, MD 20771, USA\\
}
\altaffiltext{$\dagger$}{Corresponding author (\href{mailto:\myemail}{\myemail}).}

\author{Eran~O.~Ofek\altaffilmark{51},
Mansi~M.~Kasliwal\altaffilmark{52},
Peter~E.~Nugent\altaffilmark{21,53},
Iair~Arcavi\altaffilmark{54,55},
Joshua~S.~Bloom\altaffilmark{21,53},
Shrinivas~R.~Kulkarni\altaffilmark{56},
Daniel~A.~Perley\altaffilmark{56},
Tom~Barlow\altaffilmark{56},
Assaf~Horesh\altaffilmark{51},
Avishay~Gal-Yam\altaffilmark{51},
D.~A.~Howell\altaffilmark{54,57},
and Ben~Dilday\altaffilmark{58}\\
for the PTF Collaboration
}
\affil{$^{51}$ Department of Particle Physics and Astrophysics, The Weizmann Institute of Science, Rehovot 76100, Israel\\
$^{52}$ The Observatories, Carnegie Institution for Science, 813 Santa Barbara Street, Pasadena, CA 91101, USA\\
$^{53}$ Dept.~of Astronomy, University of California, Berkeley, CA 94720, USA\\
$^{54}$ Las Cumbres Observatory Global Telescope, 6740 Cortona Drive, Suite 102, Goleta, CA 93111, USA\\
$^{55}$ Kavli Institute for Theoretical Physics, University of California, Santa Barbara, CA 93106, USA\\
$^{56}$ Cahill Center for Astrophysics, California Institute of Technology, Pasadena, CA 91125, USA\\
$^{57}$ Department of Physics, University of California, Santa Barbara, CA 93106, USA\\
$^{58}$ North Idaho College, 1000 W Garden Ave, Coeur d'Alene, ID 83814\\
}

\author{Phil~A.~Evans\altaffilmark{59},
and Jamie~A.~Kennea\altaffilmark{30}\\
for the Swift Collaboration
}
\affil{$^{59}$ Department of Physics and Astronomy, University of Leicester, Leicester, LE1 7RH, UK\\
}

\author{W.~S.~Burgett\altaffilmark{60},
K.~C.~Chambers\altaffilmark{60},
N.~Kaiser\altaffilmark{60},
C.~Waters\altaffilmark{60},
H.~Flewelling\altaffilmark{60},
J.~L.~Tonry\altaffilmark{60},
A.~Rest\altaffilmark{61},
and S.~J.~Smartt\altaffilmark{62}\\
for the Pan-STARRS1 Science Consortium
}
\affil{$^{60}$ Institute for Astronomy, University of Hawaii at Manoa, Honolulu, HI 96822, USA\\
$^{61}$ Space Telescope Science Institute, 3700 San Martin Drive, Baltimore, MD 21218, USA\\
$^{62}$ Astrophysics Research Centre, School of Mathematics and Physics, Queen's University Belfast, Belfast, BT7 1NN, UK\\
}

\begin{abstract}
  The IceCube neutrino observatory pursues a follow-up program selecting
  interesting neutrino events in real-time and issuing alerts for
  electromagnetic follow-up observations.  In March 2012, the most significant
  neutrino alert during the first three years of operation was issued by
  IceCube. In the follow-up observations performed by the Palomar Transient
  Factory (PTF), a Type IIn supernova (SN) PTF12csy was found \SI{0.2}{\degree}
  away from the neutrino alert direction, with an error radius of
  \SI{0.54}{\degree}.  It has a redshift of $z=0.0684$, corresponding to a
  luminosity distance of about \SI{300}{Mpc} and the Pan-STARRS1 survey shows
  that its explosion time was at least 158 days (in host galaxy rest frame)
  before the neutrino alert, so that a causal connection is unlikely.  The
  \emph{a posteriori} significance of the chance detection of both the
  neutrinos and the SN at any epoch is \SI{2.2}{\ensuremath{\sigma}} within
  IceCube's 2011/12 data acquisition season.  Also, a complementary neutrino
  analysis reveals no long-term signal over the course of one year. Therefore,
  we consider the SN detection coincidental and the neutrinos uncorrelated to
  the SN\@. However, the SN is unusual and interesting by itself: It is
  luminous and energetic, bearing strong resemblance to the SN IIn 2010jl, and
  shows signs of interaction of the SN ejecta with a dense circumstellar
  medium.  High-energy neutrino emission is expected in models of diffusive
  shock acceleration, but at a low, non-detectable level for this specific
  SN\@.  In this paper, we describe the SN PTF12csy and present both the
  neutrino and electromagnetic data, as well as their analysis.
\end{abstract}


\section{INTRODUCTION}
\setcounter{footnote}{0}

IceCube is a cubic-kilometer-sized neutrino detector installed in the ice at the
geographic South Pole between depths of 1450\,m and 2450\,m \citep{achterberg}.
It consists of an array of \num{5160} photon sensors, called Digital Optical Modules (DOMs), attached
to 86 cables, called strings.
Detector construction started in 2005 and finished in December 2010. Neutrino
observation relies on the optical detection of Cherenkov radiation emitted
by secondary particles produced in neutrino interactions in ice or
bedrock near IceCube.
Due to the small neutrino interaction cross-section, the kilometer-scale
detector has an effective area of only
\SIrange{0.25}{10}{m^2} for muon
neutrinos of \SIrange{1}{10}{TeV} energy \citep{IceCubePointSource}.
As part of the Optical Follow-Up (OFU) program, the IceCube neutrino
observatory records high-energy ($\sim \SI{100}{GeV} \dots
\SI{1}{PeV}$) neutrino events at a rate of about $\SI{3}{mHz}$, about
250 per day.  Those events are mostly ($\sim 90\%$) cosmic-ray
induced neutrinos from the atmosphere, referred to as atmospheric
neutrinos, with about a 10\% contamination of cosmic-ray induced
atmospheric muons.

Routine neutrino analyses in IceCube, which are referred to as
\emph{offline} analyses, are performed after a certain amount of data,
e.g.\ one or several years, has been collected.
They benefit from events being put
through computationally expensive reconstructions, as well as information on
detector performance that become available only days or weeks after data
acquisition. In contrast, neutrino analyses running \emph{online} will not
have access to such information, but have the advantage of being
near real-time---results are available with a latency of $\sim \SI{3}{min}$.
With such a short latency neutrino analysis, multi-wavelength follow-up
observations can be triggered by neutrino events. These follow-up data have the
potential to reveal the electromagnetic counterpart of a transient neutrino
source, which might otherwise be missed and thus be unavailable for further
observations.
In addition, the coincident detection of neutrino and electromagnetic emission
can be statistically more significant and provide more information about the
physics of the source than the neutrino detection alone. Another advantage of
an online analysis is the prompt availability of the reconstructed neutrino
dataset and thus the possibility of fast response analyses.  Thus, IceCube's
online neutrino analysis efforts have also enabled fast $\gamma$-ray burst
(GRB) searches like the one following GRB 130427A, published in a GCN Circular
\citep{fast-grb}.

The online search for short transient neutrino sources (order of $\leq
\SI{100}{s}$) is mostly motivated by models of neutrinos from long duration
GRBs \citep{WaxmanBahcall1997,Murase2008,Murase2006b,Becker2006} and from
choked jet supernovae (SNe) \citep{Razzaque2004,AndoBeacom2005}. The two source
classes are related: Both are thought to host a jet, which is highly
relativistic in case of long GRBs, but only mildly relativistic in case of
choked jet SNe.
Long GRB progenitors are conceived to be Wolf-Rayet stars that have
lost their outer hydrogen and helium envelope \citep{Meszaros2006},
while choked jet SNe can still have those outer layers
\citep{AndoBeacom2005} which are important for efficient high-energy
(HE) neutrino production.
The choked jet is more baryon-rich and has a
much lower Lorentz factor $\Gamma \approx 3$ than the GRB jet with
$\Gamma \gtrsim 100$. It cannot penetrate the stellar envelope and
remains optically thick, making it invisible in $\gamma$-rays.
The neutrinos produced at \si{TeV} energies can escape nevertheless and may
trigger the discovery of the SN in other channels.  In particular, for a bright
nearby SN, a neutrino detection would enable the acquisition of optical data
during the rise of the light curve, strengthening the time correlation between
the neutrino burst and the optical supernova via the improved estimate of the
explosion time \citep{Cowen2010}.
Mildly relativistic jets may occur in a much larger fraction of
core-collapse SNe than highly relativistic jets, i.e.\ GRBs
\citep{Razzaque2004,AndoBeacom2005}.

Detections of neutrinos from GRBs and choked jet SNe would be a remarkable
discovery and could provide important insight into the supernova-GRB
connection and the underlying jet physics \citep{AndoBeacom2005}.
Both sources are expected to emit a short, about \SI{10}{s} long burst
of neutrinos \citep{AndoBeacom2005} either \SIrange{10}{100}{s} before
or at the time of the $\gamma$-ray burst (if detectable)
\citep{Meszaros2006}, setting the natural time scale of the neutrino
search.
After recording the neutrino burst, follow-up observations can be used to
identify the counterpart of the transient neutrino source.
A $\gamma$-ray burst can be identified either via the prompt $\gamma$-ray emission
lasting up to about $\SI{150}{s}$ \citep{Baret2011} or via the optical and X-ray
afterglow lasting up to several hours \citep{Swift_overview}. The latter
involves instruments of limited field of view and thus requires telescope slew,
but has the advantage of much better angular resolution well below an arc
minute \citep{Swift_overview}.
A fast response within minutes to hours is required for
a GRB afterglow follow-up.
A choked jet SN is found by detecting a shock breakout or a SN light curve in
the follow-up images, slowly rising and then declining within weeks after the
neutrino burst.  Following this scientific motivation, an online neutrino
analysis,
targeted at SN and GRB afterglow detection,
was installed at IceCube in 2008 (see Section~\ref{sec:OFU}).

In addition to the transient neutrino emission within \SI{100}{s}, as
discussed above, SNe can be promising sources of high-energy
neutrinos over longer time scales.
In this paper, the class of Type IIn SNe
\citep{Schlegel1990,Filippenko1997} is explored further. These are
core-collapse SNe embedded in a dense circumstellar medium (CSM) that
was ejected in a pre-explosion phase. Following the explosion, the SN
ejecta plow through the dense CSM and collisionless shocks can form
and accelerate particles, which may create high-energy neutrinos. This
is comparable to a SN remnant, but on a much shorter time scale of
\SIrange{1}{10}{months} \citep{murase-IIn,Katz2012}.

SNe IIn (``n'' for narrow) are spectrally characterized by the
presence of strong emission lines, most notably H$\alpha$, that have a
narrow component, together with blue continuum emission
\citep{Schlegel1990,Filippenko1997}. The narrow component is interpreted to
originate from surrounding \species{H}{ii} regions, and the generally slow
spectral evolution is due to the presence of a high-density CSM
\citep{Schlegel1990}.  Since interaction of the SN ejecta with the dense CSM
can lead to the conversion of a large fraction of the ejecta's kinetic energy
to radiation, SNe IIn are on average more luminous than other SNe II
\citep{Richardson2002}. They generally fade quite slowly and some belong to the
most luminous SNe \citep{Filippenko1997}.
There is much diversity within the subclass of SNe IIn
\citep{Filippenko1997,Richardson2002}, both spectroscopically and
photometrically, which can be explained by a diversity of progenitor
stars and mass loss histories prior to explosion \citep{Moriya2011}.
Recently, there have been observations of eruptions prior to SN IIn
explosions associated with mass loss which explain the existence of
the dense CSM shells \citep[e.g.][]{Ofek2014b}.

In this paper, we report the discovery of a Type IIn SN in the
optical observations triggered by an IceCube neutrino alert from March 2012,
and analyze the available neutrino and electromagnetic observations.  The SN is
already at a late stage at the time of the neutrino detection, which means that
the neutrino-SN connection is presumably coincidental.  The paper is organized
as follows: Section~\ref{sec:OFU} introduces the optical follow-up system of
IceCube. Section~\ref{sec:alert} gives details about the neutrino alert
triggering the follow-up observations that led to the SN discovery.
Section~\ref{sec:highenergy}
reports limits from a complementary offline neutrino search and
X-ray limits. The UV and optical data that were obtained are discussed in depth
in Section~\ref{sec:lowenergy}. We finally summarize the results and give a
conclusion in Section~\ref{sec:conclusion}.

\section{THE OPTICAL AND X-RAY FOLLOW-UP SYSTEM}
\label{sec:OFU}

In late 2008, an online neutrino event selection was set up at IceCube, looking
for muons produced by charged current interactions of muon neutrinos in or near
the IceCube detector.  The analysis is running in real-time within the limited
computing resources at the South Pole, capable of reconstructing and filtering
the neutrinos and sending alerts to follow-up instruments with a latency of
only a few minutes \citep{OFU-ICRC,OFUpaper,mohr}.

The optical (OFU) and X-ray (XFU) real-time follow-up programs currently
encompass three follow-up instruments: the Robotic Optical Transient
Search Experiment (ROTSE) \citep{ROTSE_overview}, the Palomar
Transient Factory (PTF) \citep{PTF_overview,PTF_science} and the Swift
satellite \citep{Swift_overview}. These triggered observations were
supplemented with a retrospective search through the Pan-STARRS1 3$\pi$ survey data
\citep{Pan-STARRS,magnier2013}, which is discussed further in
Section~\ref{sec:lowenergy}.
In addition, there is also a real-time $\gamma$-ray follow-up program (GFU)
targeting slower transients (time scale of weeks), e.g.\ flaring Active
Galactic Nuclei, that is sending alerts to the $\gamma$-ray telescopes MAGIC
and VERITAS \citep{OFU-ICRC}.

The background of cosmic-ray induced muons from the atmosphere
above the detector amounts to $\sim 10^6$ muon events per neutrino event.
In a first step, it is reduced by limiting the sample to the northern
hemisphere, using the Earth as a muon shield and selecting only muon tracks
that are reconstructed as up-going in the detector. Afterwards, cuts on quality
parameters similar to those in \citet{IceCubePointSource} are applied to reject
mis-reconstructed muon events:
After an initial removal of likely noise signals, a chain of reconstructions is
performed, which utilizes the spatial and temporal distribution of recorded
photons on the DOMs. Starting with a simple linear track algorithm, more
advanced reconstructions are employed, seeded with the respective preceding
reconstruction. The advanced reconstructions maximize a likelihood that
accounts for the optical properties of the ice for photon propagation
\citep{IceCubeLikelihoodReco}.  The final reconstruction in this chain is used
to select events that are up-going in the detector.
At this point, because of the vast amount of atmospheric muons,
the data are still dominated by (down-going) muons that are mis-reconstructed
as up-going.
To reject this remaining background, high quality events are selected,
where the selection parameters are derived from the value of the maximized
track likelihood
and from the number and geometry of recorded signals with a detection time
compatible with unscattered photon propagation.  Alternatively, events with a
large number of total recorded photon signals are selected.
The resulting analysis sample has an event rate of about $\SI{3}{mHz}$, of which
$\sim 90\%$ are atmospheric muon neutrinos that have passed through the Earth
and $\sim 10\%$ are contamination of cosmic-ray induced muons from the
atmosphere.

In the analysis sample, the median angular resolution of the neutrino
direction is about \SI{1}{\degree} for a multi-TeV muon neutrino
charged current interaction event,
and \SI{0.6}{\degree} or less for \SI{100}{TeV} and higher energies. The
angular resolution of the sample is estimated using Monte Carlo (MC)
simulation. Additionally, an estimator of the directional
uncertainty is computed for each event, which is based on the shape of
the reconstruction likelihood close to the found maximum
\citep{Neunhoeffer}. It is calibrated such that its median matches the
MC derived angular resolution. We define the \emph{bulk} of a sample
as events within the central 90\% of the energy distribution. The bulk
of the main background contained in the analysis sample, atmospheric
neutrinos, has energies between \SI{160}{GeV} and \SI{7}{TeV}. In
contrast, the bulk of signal events from an unbroken
$E_\nu^{-2}$ power-law spectrum would have energies $\SI{1.2}{TeV}
\lesssim E_\nu \lesssim \SI{1.2}{PeV}$. Signal neutrinos from
GRBs are expected to follow a spectrum similar to $E_\nu^{-2}$,
which has cut-off energies between $\sim \SI{1}{PeV}$ to $\sim \SI{1}{EeV}$
\citep{Murase2006a}.
Choked jet SNe are predicted to have lower cut-off energies around \SI{20}{TeV}
\citep{AndoBeacom2005} so that the overlap with atmospheric neutrinos is
presumably much larger, while IIn neutrinos may have higher cut-offs of
\SIrange{70}{200}{TeV} (see Section~\ref{sec:offline}).

In order to suppress background from atmospheric neutrinos, a multiplet of at least two
neutrinos within 100 seconds and angular separation of \SI{3.5}{\degree} or
less is required to trigger an alert. In addition, since 2011 mid-September, a
test statistic is used, providing a single parameter for selection of the
most significant alerts.
It was derived as the analytic maximization of a likelihood ratio following
\citet{IceCube-PS-timedep}, for the special case of a neutrino doublet with
rich signal content:
\begin{align}
    \begin{split}
    \lambda &= \frac{\Delta\Psi^2}{\sigma_q^2} + 2 \ln(2 \pi \sigma_q^2) \\
    &- 2 \ln\left( 1 - \exp\left(-\frac{\theta_A^2}{2\sigma_w^2}\right) \right)
            + 2 \ln\left( \frac{\Delta T}{\SI{100}{s}} \right)
    \end{split}
  \label{eq:alert-llh}
\end{align}
where the time between the neutrinos in the doublet is denoted as $\Delta T$,
and their angular separation as $\Delta\Psi$. The quantities $\sigma_q^2 =
\sigma_1^2 + \sigma_2^2$ and $\sigma_w^2 = \left(1/\sigma_1^2 +
1/\sigma_2^2\right)^{-1}$ depend on the event-by-event directional
uncertainties $\sigma_1$ and $\sigma_2$ of the two neutrino events, typically
$\sim \SI{1}{\degree}$.
The angle $\theta_A$ corresponds to the circularized angular radius of the
field of view (FoV) of the follow-up telescope. It is set to
$\SI{0.5}{\degree}$ for Swift and $\SI{0.9}{\degree}$ for ROTSE and
PTF\@.

The test statistic $\lambda$ is smaller for more signal-like alerts, which have small
separation $\Delta\Psi$, small time difference $\Delta T$ and a high
chance to lie in the FoV of the telescope. Thus, $\lambda$ is a useful
parameter to separate signal and background alerts. For each follow-up
program, a specific cut on $\lambda$ is applied in order to send the
most significant alerts to the follow-up instruments.
In the 2011/12 data acquisition (DAQ) season, which is discussed here, a cut
of $\lambda<\num{-7.4}$ was used for the ROTSE follow-up, while cut values of
\num{-10.3} and \num{-8.8} were used for PTF and Swift.
Multiplets of multiplicity higher than two are passed directly to all
follow-up instruments. Since the expected background rate is low
($\sim 0.03$ per year), each observation of a triplet or higher order multiplet is
significant by itself.

\textbf{ROTSE} \citep{ROTSE_overview} is a network of four optical telescopes with
\SI{0.45}{m} aperture and $\SI{1.85}{\degree} \times \SI{1.85}{\degree}$ FoV,
located in Australia, Texas, Namibia and Turkey. Since late 2012, only
the two northern hemisphere telescopes continue operation. ROTSE is a
completely automatic and autonomous system that can receive alerts,
perform observations and send resulting data without requiring human
interaction. The limiting magnitude of $\sim$~\SIrange{16}{17}{mag} is
however insufficient to discover faint or far SNe.
For instance, for a very bright SN with \SI{-20}{mag} absolute
magnitude, the detection radius is about \SIrange{160}{250}{Mpc},
while a faint SN with \SI{-17}{mag} is only visible within a radius of
\SIrange{40}{65}{Mpc}.
IceCube has been sending $\sim 25$ alerts per year to ROTSE, since 2008
December.
The first 116 alerts, with a background expectation of $104.7\pm10.2$ alerts,
were followed up with a median latency of \num{27.2} hours between the neutrino
alert and start of the first follow-up observation.

\textbf{PTF} \citep{PTF_overview,PTF_science} is a survey based at the Palomar
Observatory in California, USA\@. It utilizes the $\SI{1.2}{m}$ Oschin Schmidt
telescope on Mount Palomar. The focal plane is equipped with a mosaic of 11
CCDs with field of view of \SI{7.26}{deg^2}. The typical $R$-band limiting
magnitude of PTF during dark time is about \SI{21}{mag}. All the PTF data are
reduced using the LBNL real-time pipeline responsible for transient
identification and the IPAC pipeline described in \citet{Laher2014}. The image
photometric calibration is described in \citet{Ofek2012}. PTF pursues a number
of science goals, most notably the discovery and observation of SNe. Several
other telescopes in Palomar and at other locations
can be used for photometric and spectroscopic follow-up observation.
IceCube has been sending $\sim 7$ alerts per year to PTF, since August 2010.
The first 23 alerts, with a background expectation of $19.2\pm4.4$ alerts, were
followed up with a median latency of \num{34.9} hours between the neutrino
alert and start of the first follow-up observation.

\textbf{Swift} \citep{Swift_overview} is a satellite operated by NASA and
boards various instruments: a \SIrange{170}{600}{nm} ultraviolet/optical
telescope (UVOT), a \SIrange{0.3}{10}{keV} X-ray telescope (XRT) and a
\SIrange{15}{150}{keV} hard X-ray Burst Alert Telescope (BAT). Swift's main
goal is the discovery and study of GRBs, of which it detects about \num{100}
per year \citep{Lien2014} (one third of all
GRBs)\footnote{\url{http://gcn.gsfc.nasa.gov/} and \url{http://grbweb.icecube.wisc.edu}}.
IceCube's X-ray follow-up program triggers Swift's XRT, which can provide
valuable information by observing a GRB afterglow in X-rays. The XRT has a FoV
of only $\SI{0.4}{\degree}$ in diameter, hence Swift performs seven pointings
for each IceCube follow-up, resulting in an effective FoV of about
\SI{1}{\degree} in diameter. IceCube has been sending $\sim 6$ alerts per year
to Swift, since February 2011.
The first 18 alerts, with a background expectation of $18.0\pm4.2$ alerts, were
followed up with a median latency of \num{1.9} hours between the neutrino alert
and start of the first follow-up observation \citep[see][]{EvansSwift}.

\section{NEUTRINO ALERT AND DISCOVERY OF PTF12CSY}
\label{sec:alert}

On 2012 March 30 (MJD 56016), the most significant alert since initiation of
the follow-up program (significance of $\sim \SI{2.7}{\ensuremath{\sigma}}$,
converting the $\lambda$ cumulative distribution function value to single-sided Gaussian std.\ deviations)
was recorded and sent to ROTSE and PTF simultaneously. The significance was
also above the threshold for Swift ($\sim \SI{1}{\ensuremath{\sigma}}$),
however it was within Swift's moon proximity constraint\footnote{Swift is
unable to observe sources closer than \SI{15}{\degree} to the moon.}, which
delayed the observations by three weeks.
The two neutrino events causing the alert happened on 2012 March 30 at
01:06:58 UT (MJD 56016.046505) and 1.79 seconds later,
with an angular separation of
\SI{1.32}{\degree}. The combined average neutrino direction is at right ascension
\si{6^h57^m45^s}
and declination \si{17\degree11'24''}
in J2000 with an error radius of $\sigma_w = \SI{0.54}{\degree}$.
This average is a weighted arithmetic mean, weighting the individual directions
with their inverse squared error, given by the event-by-event directional
uncertainty (s.a.). The error $\sigma_w$ is defined after Eq.~\ref{eq:alert-llh}.
This assumes that the two neutrino events were emitted by a point source at a
single fixed position. A variance of the individual true neutrino directions
does not need to be taken into account, since the assumed intrinsic variance is
zero in case of a point source. This leads to a relatively small error on the
average direction.

The main event properties are summarized in Table~\ref{tab:properties}: the
occurrence time on 2012 March 30, the reconstructed muon energy proxy
$\hat{E}_\mu$ \citep[see][]{IceCubeEnergyReco}, and the estimated directional
error $\sigma_\Psi$. The quantity $\hat{E}_\mu$ is a fit parameter and serves as
a proxy for the muon energy, however it is not an estimator of the true muon
energy.
The energy $E_\nu$ of the neutrino that produced the muon is not directly
observable, since only the muon crossing the detector is accessible. However,
using Monte Carlo (MC) simulated neutrino events, one can use the muon energy
proxy $\hat{E}_\mu$ to compare with MC events having a similar $\hat{E}_\mu$
value. From those MC events, a distribution of the true muon energy $E_\nu$ can
be derived, which depends on the assumed underlying neutrino energy spectrum.
$E_\nu$ probability density functions are filled with true energy values of MC
events that have the reconstructed muon energy proxy $\hat{E}_\mu$ not more than
10\% and the reconstructed zenith angle $\cos(\theta)$ not more than 0.1 away
from the observed values of the two alert events.  The median and the central
90\% C.L.\ interval are calculated from the $E_\nu$ probability density
functions.  The results are listed in Table~\ref{tab:properties}, where the
assumed neutrino spectrum is given in parentheses.

\begin{deluxetable}{cccccc}
    \tablewidth{\columnwidth}
    \tablecaption{Properties of the neutrino alert events\label{tab:properties}}
    \tablehead{\colhead{\parbox{0.8cm}{\centering Time (\si{UT})}} &
        \colhead{\parbox{0.8cm}{\centering $\sigma_\Psi$ (\si{\degree})}} &
        \colhead{\parbox{0.8cm}{\centering $\hat{E}_\mu$\tablenotemark{a} (\si{GeV})}} &
        \colhead{\parbox{1.6cm}{\centering $E_\nu$\tablenotemark{b} (Atm.)  (\si{TeV})}} &
        \colhead{\parbox{1.5cm}{\centering $E_\nu$\tablenotemark{b} ($E^{-3}$) (\si{TeV})}} &
        \colhead{\parbox{1.5cm}{\centering $E_\nu$\tablenotemark{b} ($E^{-2}$) (\si{TeV})}}}
    \startdata
        01:06:58 & 0.96           & \num{1155}    & $0.5_{-0.4}^{+2.9}$ & $0.7_{-0.5}^{+5.6}$  & $5.4_{-5.0}^{+292.0}$   \\
        01:07:00 & 0.66           & \num{3345}    & $0.9_{-0.7}^{+6.7}$ & $1.5_{-1.3}^{+14.8}$ & $15.7_{-14.5}^{+611.5}$
    \enddata
    \tablenotetext{a}{$\hat{E}_\mu$ is only a
    proxy correlated with muon energy, but not an estimator of the true muon energy.}
    \tablenotetext{b}{$E_\nu$ is median neutrino energy with 90\% C.L.\ error interval.}
\end{deluxetable}

\begin{figure}[tbp]
    \plotone{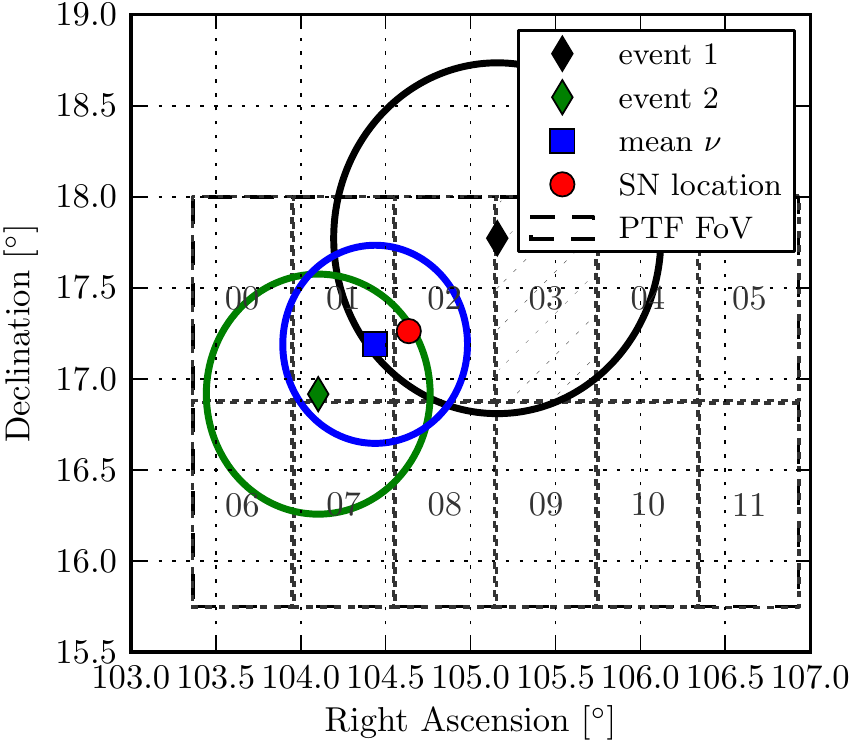}
    \caption{Map of the sky with the two neutrino event directions,
    the average neutrino direction, and the location of SN PTF12csy.
    Estimated reconstruction errors are indicated with circles, the
    PTF FoV is shown as dashed box. The positions of the PTF survey
    camera CCD chips are plotted with dotted lines and the chip number
    is printed on each chip's field (cf.\ \citet{PTF_overview}). Note
    that chip 03 is not operational and thus hatched in the plot.}
    \label{fig:map}
\end{figure}

\begin{figure}[tbp]
    \plotone{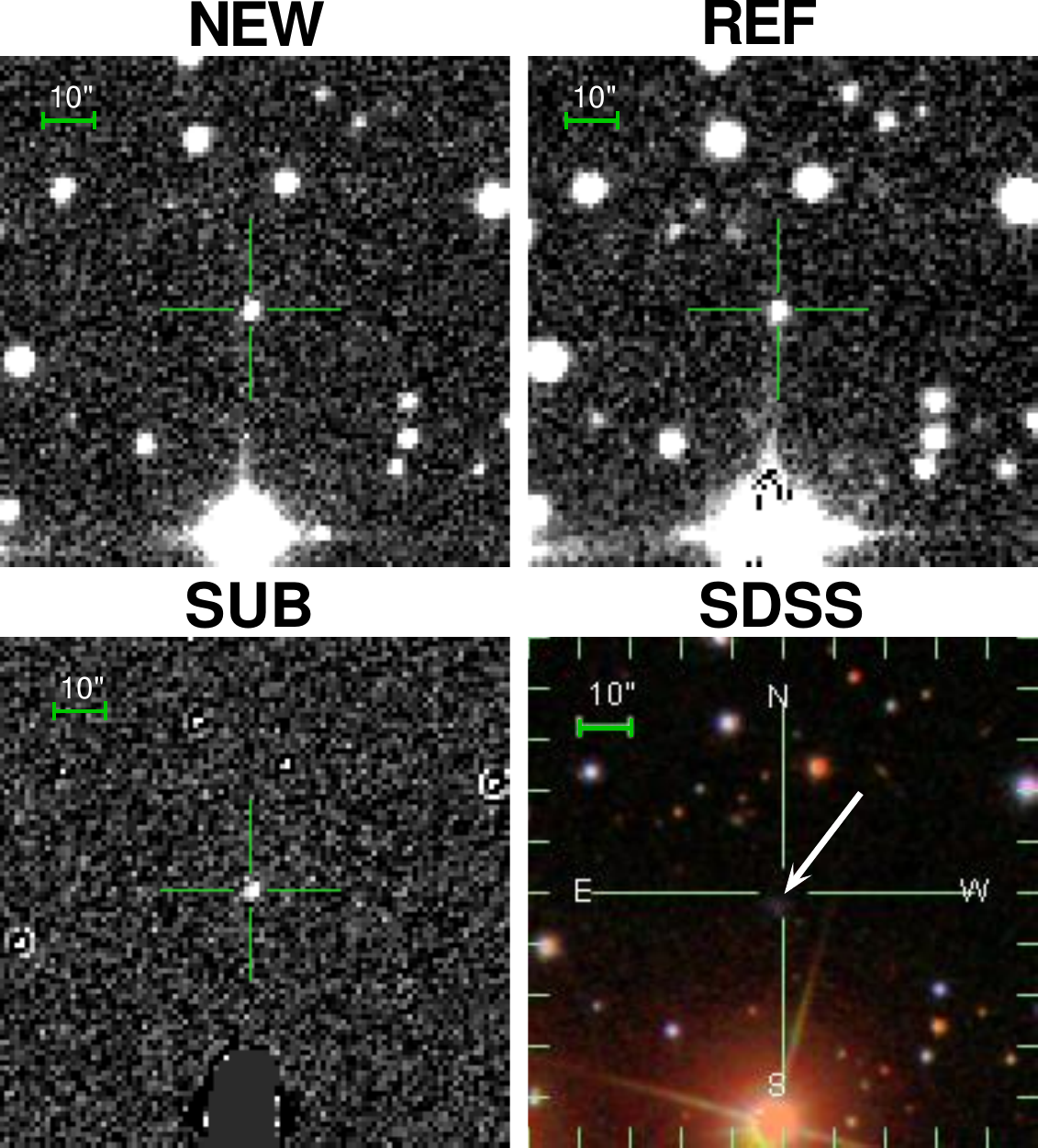}
    \caption{New image, reference image and post-subtraction image of
    the PTF discovery of PTF12csy from 2012 April 09, with the location
    of PTF12csy in the center. This image shows only a small fraction
    of the PTF FoV. The image from the Sloan Digital Sky Survey (SDSS-III)
    DR12 \citep{Eisenstein2011,Gunn2006,Ahn2014,Alam2015}
    is shown for reference, showing a faint host galaxy to the south of the
    SN.}
    \label{fig:subtraction}
\end{figure}

Follow-up observations at the direction of the neutrino alert were
performed with multiple instruments (see Section~\ref{sec:obs}).
In the PTF images, a core-collapse supernova, named \emph{PTF12csy},
was discovered at right ascension
\si{6^h58^m32^s{.}744}
and declination
\si{17\degree15'44{.}37''} (J2000), only \SI{0.2}{\degree} away from the
average neutrino direction, see Figs.~\ref{fig:map} and~\ref{fig:subtraction}.
This was a promising candidate for the source of the neutrinos, but a search of the
Pan-STARRS1 archive (see Section~\ref{sec:obs}) revealed that it was already at
least 169 observer frame days old, i.e.\ 158 days in host galaxy rest frame, at
the time of the neutrino alert.
Therefore, it is highly unlikely that the neutrinos were produced by a jet at
the SN site, as this is expected to happen immediately after core collapse in
the choked jet scenario \citep{AndoBeacom2005}.

However, steady neutrino emission on a time scale of several months is
a possibility and explored in Section~\ref{sec:offline}.

\subsection{Significance of Alert and SN Detection}

The value of the test statistic $\lambda$ for the neutrino doublet
amounts to $-18.1$.
The background distribution of $\lambda$ is constructed from experimental data,
containing mostly atmospheric neutrinos, by randomly permuting (shuffling) the
event times and calculating equatorial coordinates, i.e.\ right ascension and
declination, from local coordinates, i.e.\ zenith and azimuth angle, using the
new times. That way, all detector effects are entirely preserved, e.g.\ the
distribution of the azimuth angle, which has more events at angles where
detector strings are aligned, and the time distribution, which is affected by
seasonal variations.  At the same time, all potential correlations between the
events in time and space, and thus a potential signal, are destroyed.

The false alarm rate (FAR) for an alert with $\lambda \leq -18.1$
is \SI{0.226}{yr^{-1}}, calculated via integration of the $\lambda$
distribution below \num{-18.1}.
Considering the OFU live time of 220.1 days in the data acquisition season of
the alert, September 2011 to May 2012,
yields $N(\lambda < -18.1) = 0.136$ false alerts. Hence,
the probability, or p-value, for one or more alerts
at least as signal-like to happen by chance
in this period is $1 - P_\mathrm{Poisson}(0;N(\lambda < -18.1)) \approx 12.7\%$.
The OFU system had already been sending alerts to PTF for $\sim 460$ days at
the time of the alert. Scaling up the number of expected alerts with $\lambda
\leq -18.1$, one derives a probability of $\sim 24\%$ during 460 days.

The estimated explosion time of SN PTF12csy does not fall within the \emph{a
priori} defined time window for a neutrino-SN coincidence of ${\cal
O}(\SI{1}{day})$. It is thus not considered an a priori detection of the
follow-up program. Despite this fact, for illustrative purposes, we calculate
the \emph{a posteriori} probability that a random core-collapse SN (CCSN) of
any type, at any stage after explosion, is found coincidentally within the
error radius of this neutrino doublet and within the luminosity distance of
PTF12csy, i.e.\ $\SI{300}{Mpc}$. The number of such random SN detections is
\begin{equation}
    \overline{N}_\mathrm{det} =
    \frac{\Omega_{s}}{4 \pi} \int_0^{\SI{300}{Mpc}}
    \frac{dN_\mathrm{SN}}{dt \, dV} \; T(m_\mathrm{lim}, \hat{M}, r) \; 4 \pi r^2 \, dr
  \label{SN-det-prob}
\end{equation}
where $\Omega_{s}$ is the solid angle of the doublet error circle (blue circle in Fig.~\ref{fig:map}),
which is $\sim \SI{0.93}{(\degree)^2}$.
For the volumetric CCSN rate $dN_\mathrm{SN}/(dt \, dV)$, a value
of $\SI{0.78e-4}{Mpc^{-3}.yr^{-1}}$ is used, \citet[see][sec.~4.1]{Horiuchi}.
The control time $T(m_\mathrm{lim}, \hat{M}, r)$ is the average time window in
which a SN is detectable, i.e.\ brighter than the limiting magnitude.  It
depends on the distance to the source $r$, the peak absolute magnitude
$\hat{M}$ of the SN, the limiting magnitude $m_\mathrm{lim}$ of the telescope,
and the shape of the light curve which is adopted from a SN template
web page by P.~E.~Nugent\footnote{\url{https://c3.lbl.gov/nugent/nugent_templates.html}}.
It is assumed that $\hat{M}$ follows a Gaussian distribution with mean
$\SI{-17.5}{mag}$ and standard deviation $\sigma=\SI{1}{mag}$, based on
\citet{Richardson2002}. For PTF, $m_\mathrm{lim} = \SI{19.5\pm1}{mag}$ is
assumed.

The resulting expectation value for coincidental SN detections is
$\overline{N}_\mathrm{det} \approx 0.016$, which results in a Poisson
probability of $\sim 1.6\%$ to detect any CCSN within the neutrino alert's
error radius.  Combining this probability with the probability of $12.7\%$ for
the neutrino alert, Fisher's method \citep{Brown1975,LittellFolks1971,
FisherBook1950,FisherBook1990} delivers a combined p-value of $1.4\%$, corresponding to a
significance of \SI{2.2}{\ensuremath{\sigma}}. For the total live time of 460 days, the
combined p-value is $2.5\%$ which corresponds to a \SI{2.0}{\ensuremath{\sigma}}
significance. This means,
even ignoring the a posteriori nature of the p-value,
a chance coincidence of the neutrino doublet and the SN detection
cannot be ruled out
and thus we consider the SN detection to be coincidental.

The following section reports about the available high-energy follow-up data.
Limits on a possible long-term neutrino emission from PTF12csy are set using
one year of IceCube data. Limits on the X-ray flux were obtained using the
Swift satellite. Section~\ref{sec:lowenergy} deals with the analysis of the
low-energy optical and UV data as the SN detection is significant and
interesting by itself.

\section{HIGH-ENERGY FOLLOW-UP DATA}
\label{sec:highenergy}

\subsection{Offline Analysis of Neutrino Data}
\label{sec:offline}

Type IIn supernovae, such as PTF12csy, are a promising class of high-energy
transients (see \citet{murase-IIn}). The expected duration of neutrino emission
from SNe IIn is \numrange{1}{10} months, hence it is extremely unlikely that
two neutrinos arrive within less than \SI{2}{s}, so late after the SN
explosion. However, to test the possibility of a long-term emission, a search
for neutrinos from PTF12csy within a search window of roughly one year is
conducted.

After the core-collapse of a SN IIn, the supernova ejecta are crashing into
massive circumstellar medium (CSM) shells, producing
a pair of shocks: a forward and a reverse shock.
Cosmic rays (CRs) may be accelerated and multi-TeV neutrinos produced,
potentially detectable with IceCube. The collisionless shocks generating the
neutrinos are expected to generate X-rays as well at late times (see e.g.\
\citet{Katz2012,Svirski,Ofek2013}), but no X-rays were detected for PTF12csy,
likely because of the large distance to the SN.

Following \citet{murase-IIn} and \citet{Murase2014} \citep[see also][]{Katz2012},
we model high-energy neutrino emission from PTF12csy.
As a simplified approach, in order to get an order-of-magnitude estimate of the
expected event rate, we perform the following calculation:
The CSM density profile is calculated using \citet[eq.~4, eq.~A4]{Murase2014}.
The proton spectrum is modeled with a power law index \num{-2}, as in
\citet[eq.~A7]{Murase2014}, with a cut-off energy given by
the maximum energy of accelerated protons. The latter is determined by comparing the
proton acceleration time scale either with the dynamical time scale
\citep[eq.~28]{Murase2014}, or, if $pp$ energy losses are relevant, with the cooling
time scale \citep[eq.~30]{Murase2014}. The lower of the two maximum energies
is the one that needs to be considered.
The proton spectrum is normalized to the total CR energy $E_\text{CR}$ by
assuming that a fraction of the kinetic energy of the ejecta $E_\text{ej}$ is
converted into CRs,
that is $E_\text{CR} = \epsilon_\text{CR} E_\text{ej}$ \citep[compare][eq.~3,
eq.~25, using CSM shell mass $M_\text{CSM} \gg $ SN ejecta mass
$M_\text{ej}$]{Murase2014}.
Other model parameters are the break-out radius $R_\text{bo}$ and shock
velocity $v_\text{shock}$. The expected neutrino spectrum from $pp$ interaction
is derived from the semi-analytical description in \citet{Kelner2006}, taking
into account the meson production efficiency \citep[eq.~35, eq.~36]{Murase2014}.
It is distance-scaled and folded with IceCube's effective area from
\citet{IceCubePointSource} to obtain the expected number of events.
It is found that inserting commonly assumed values \citep{Murase2014,Margutti_2009ip,murase-IIn}
of $E_\text{ej} = 10 E_\text{bol} = \SI{2.1e+51}{erg}$ (with
the bolometric energy found in Section~\ref{sec:sed}),
$\epsilon_\text{CR} = 0.1$, $M_\text{ej} = \SI{10}{\solarmass}$,
$R_\text{bo} \approx \SI{6e15}{cm}$,
and $v_\text{shock} \approx \SI{5000}{km.\second^{-1}}$, on average only
\num{0.07} IceCube neutrino detections are expected.

Despite the low expectation for the neutrino fluence, we search for
a long-term neutrino signal from PTF12csy in the IceCube data.
As a more elaborate approach compared to the simplified approximation described
above, we test the neutrino emission models A and B given in Fig.~1 of
\citet{murase-IIn}, which are two representative cases of CR
accelerating scenarios: Model A corresponds to a CSM shell with a high density
of \SI{1e11}{cm^{-3}} at a small radius of \SI{1e15.5}{cm}, while model
B is the opposite with a density of \SI{1e7.5}{cm^{-3}} at radius \SI{1e16.5}{cm}.
For the ejecta, a kinetic energy of \SI{1e51}{erg}, a velocity of
\SI{1e4}{km.s^{-1}}, and a mass of several solar masses, lower than the CSM
mass, are assumed. Model A is close to a scenario explaining superluminous SNe
IIn such as SN 2006gy, while model B is a good description for dimmer, but
longer lasting SNe like SN 2008iy. Both models have a neutrino energy spectrum
close to $E^{-2}$, with a cut-off energy around \SI{70}{TeV} for model A,
around \SI{84}{TeV} for the forward shock (FS) in model B, and around
\SI{275}{TeV} for the reverse shock (RS) in model B. In model A, only the
reverse shock is of importance for CR acceleration.  The suggested emission
time scales are $\SI{1e7}{s}=\SI{115}{d}$ for model A and
$\SI{1e7.8}{s}=\SI{366}{d}$ for model B \citep{murase-IIn}.

\begin{figure}[tbp]
    \plotone{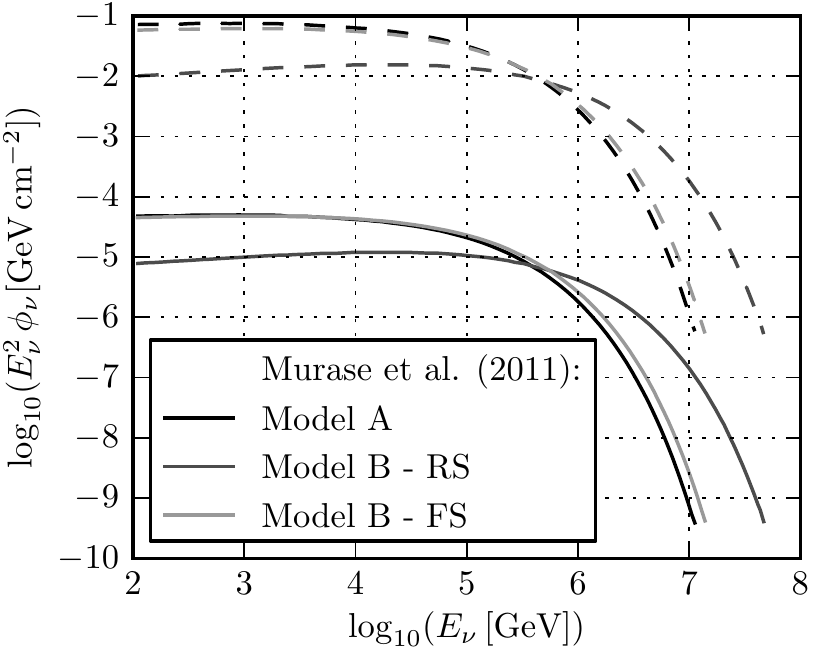}
    \caption{Neutrino fluence at Earth from PTF12csy (solid lines) and derived
    upper limits set by IceCube (dashed lines, corresponding gray scales) as
    function of energy for the tested models A, B reverse shock (RS), and B
    forward shock (FS) from \citet{murase-IIn}.}
    \label{fig:limits}
\end{figure}

Figure~\ref{fig:limits} shows the model fluences scaled down to a
luminosity distance of \SI{308}{Mpc}.
We analyze about one year of IceCube data: the entire IceCube 86
strings data acquisition season 2011/12, from 2011 May 13 to 2012
May 15. The long search window is motivated by the large uncertainty
on the explosion date (between 2011 March 21 and 2011 October 13) as
well as the long duration of neutrino emission for some scenarios like
$\sim 700$ days for model B in \citet{murase-IIn}.
For simplicity,
we assume that the entire fluence was emitted during the \SI{1}{yr}
search window. We use the neutrino sample of the IceCube optical
follow-up system and perform a statistical point source analysis of
neutrino events close to the position of the supernova, based on
\citet{IceCube-PS}.

Each neutrino candidate event $i$ is given both a signal
and background probability $S_i$ and $B_i$ which are combined in the
likelihood function
\begin{equation}
    {\cal L}(n_s) = \prod_{i=1}^N \frac{n_s}{N} S_i +
                      \left( 1 - \frac{n_s}{N} \right) B_i.
    \label{eq:likelihood}
\end{equation}
The variable $n_s$ is the number of signal events contained in the sample, which is
fitted to maximize the likelihood, and $N$ is the total
number of selected neutrino candidate events. The signal probability $S_i$ is the product of the
spatial Gaussian probability density function (PDF) and the energy PDF
$P(E_i|\phi)$, i.e.\ the probability of a signal event having reconstructed energy
$E_i$ given the neutrino spectrum $\phi$ of the source (derived from Monte
Carlo simulation):
\begin{equation}
    S_i(\sigma_i, \boldsymbol{x}_i, \boldsymbol{x}_s, E_i) =
                \frac{1}{2 \pi \sigma_i^2}
                    \exp\left(\frac{-|\boldsymbol{x}_i - \boldsymbol{x}_s|^2}{2\sigma_i^2}\right)
                        P(E_i|\phi).
    \label{eq:signal_pdf}
\end{equation}
Here, $\sigma_i$ is the event's angular error estimate, $\boldsymbol{x}_i$ is
the event's reconstructed direction, and $\boldsymbol{x}_s$ the SN or
neutrino source position.
The background probability $B_i$ contains the energy PDF
and a normalization constant for the background from atmospheric neutrinos.

We define the test statistic as likelihood ratio \citep{IceCube-PS}
\begin{equation}
    \lambda = -2 \log\left[ \frac{{\cal L}(0)}{{\cal L}(\hat{n}_s)} \right],
    \label{eq:teststat}
\end{equation}
serving as a powerful test for separating the null hypothesis from the
hypothesis of signal event contribution. Here, $\hat{n}_s$ is the number of
signal events that maximizes the likelihood and corresponds to the
most likely description of the data.

The result of the maximum likelihood fit is $\hat{n}_s = 0$ both for
model A and B, i.e.\ we see no sign of signal contribution in our
neutrino event sample. We set 90\% C.L. Neyman upper limits
(see \citet{Neyman1937}, reprinted in \citet{NeymanBook1967}) on the
tested neutrino fluence models, which amount to
$\sim 1500$ and $\sim 1300$
times the fluences given for model A and B above, respectively.
The limits are much higher than the fluence prediction because of IceCube
being insensitive to SNe IIn at such large distances.
Figure~\ref{fig:limits} shows a plot of the tested neutrino fluence and
the limits set using \SI{1}{yr} of IceCube data.

This null result and the large distance to the SN further support our
conclusion that the SN detection was coincidental.

Nevertheless, it is interesting to roughly estimate the hypothetical emitted neutrino
fluence: We take the median neutrino energies of the two alert neutrinos
from Table~\ref{tab:properties} and look
up the effective areas for the OFU neutrino sample at the respective
energies. From this, we derive a hypothetical neutrino fluence of
\SI{3.2e-4}{erg.cm^{-2}} for a source spectrum $\propto E^{-2}$
($E_\nu = \SI{5.4}{TeV}, \SI{15.7}{TeV}$), or
\SI{10.8e-4}{erg.cm^{-2}} for a source spectrum $\propto E^{-3}$
($E_\nu = \SI{0.7}{TeV}, \SI{1.5}{TeV}$).
Assuming that this neutrino fluence was emitted by
the SN, this would imply a radiated neutrino energy of $\sim
\SI{3.4e51}{erg}$ or $\sim \SI{1.2e52}{erg}$ using the luminosity
distance of $\sim \SI{300}{Mpc}$, corresponding to about 15 or
about 50 times the radiated electromagnetic energy of
$E_\mathrm{bol}=\SI{2.1e50}{erg}$ (see Section~\ref{sec:sed}). This is
higher than what can be expected, since with reasonable assumptions
that the explosion energy $E_\mathrm{ej} \leq 10 E_\mathrm{bol}$
\citep{Murase2014,Margutti_2009ip} and a fraction $\epsilon_\mathrm{CR} \leq
0.1$ of it going into CRs \citep{murase-IIn}, the energy in neutrinos should be on
the same order or less than $E_\mathrm{bol}$. Thus, also with a simple
energetic argument, isotropic neutrino emission from PTF12csy causing the
neutrino alert is implausible, especially on a time scale of seconds.  However,
a beamed emission from a jet with a small opening angle of $<\SI{30}{\degree}$
would in principle be possible.

\subsection{X-ray Observations of PTF12csy}
\label{sec:x-ray}

The Swift satellite observed the supernova four times, on 2012 April 20 (MJD
56037) and around
2012 November 15 (MJD 56246) (see Table~\ref{tab:xray}).
We perform source detection using the software developed for the 1SXPS
catalog \citep{Evans2014} on each of the four observations, and
on a summed image made by combining all the datasets. No counterpart to
PTF12csy is detected. Upper limits are generated for each of these images,
following \citet{Evans2014}.
A \SI{28}{\arcsec} radius circle centered on the optical position of
PTF12csy is used to measure the detected X-ray counts $c$ at this location, and the
expected number of background counts $c_\text{bg}$, predicted by the background
map created in the source detection process. We then use the Bayesian method of
\citet{Kraft1991} to calculate the \SI{3}{\ensuremath{\sigma}} upper limit $c_\text{UL}$
on the X-ray count rate of PTF12csy, using the XRT exposure map to correct for
any flux losses due to bad pixels on the XRT detector, and the finite size of
the circular region.

The upper limit count rate is converted to unabsorbed flux $\Phi_\text{UL}$
using the HEASARC Tool
WebPIMMS\footnote{\url{https://heasarc.gsfc.nasa.gov/Tools/w3pimms.html}},
assuming a black body model with $T=\SI{0.6}{keV}$ as in \citet{Miller_2008iy},
a Galactic hydrogen column density of \SI{1.31e21}{cm^{-2}}
\citep{Willingale}\footnote{\url{http://www.swift.ac.uk/analysis/nhtot/index.php}}
and a redshift of $z=0.0684$.
The result is a \SIrange{0.2}{10}{keV} X-ray flux $<
\SI{4.6e-14}{erg.cm^{-2}.\second^{-1}}$ for the most constraining upper limits,
corresponding to a \SIrange{0.2}{10}{keV} X-ray luminosity of $L_X <
\SI{5.2e41}{erg.\second^{-1}}$ with a luminosity distance of about
\SI{308}{Mpc}.
Using a power-law $\propto E^{-2}$ instead of a black body as an alternative
X-ray emission model, the unabsorbed flux upper limit becomes $<
\SI{7.4E-14}{erg.cm^{-2}.\second^{-1}}$, and hence, $L_X <
\SI{8.4e41}{erg.\second^{-1}}$.

\begin{deluxetable}{cccccc}
    \tablecaption{Swift XRT observations of PTF12csy\label{tab:xray}}
    \tablehead{\colhead{Time (\si{MJD})} & \colhead{Exposure (\si{\kilo\second})} &
        \colhead{$c$} & \colhead{$c_\text{bg}$} & \colhead{$c_\text{UL}$} &
        \colhead{$\Phi_\text{UL}$}}
    \startdata
        56037.15 & 4.9                 & 1   & 1.47          & 1.3           & \num{4.6}        \\
        56245.29 & 2.0                 & 1   & 0.62          & 2.1           & \num{7.4}        \\
        56246.04 & 1.2                 & 2   & 0.44          & 9.8           & \num{30.}        \\
        56247.62 & 5.0                 & 2   & 1.35          & 2.1           & \num{7.4}        \\
        Sum      & 13.0                & 6   & 3.71          & 1.3           & \num{4.6}
    \enddata
    \tablecomments{Energy range: \SIrange{0.2}{10}{keV}. $c$: measured counts
    within a \SI{28}{\arcsec} aperture. $c_\text{bg}$: expected background
    counts within the same aperture. $c_\text{UL}$: \SI{3}{\ensuremath{\sigma}} upper limit on the X-ray
    count rate in $\si{10^{-3}.s^{-1}}$. $\Phi_\text{UL}$: upper limit on the
    unabsorbed source flux in $\si{10^{-14}.erg.cm^{-2}.\second^{-1}}$.}
\end{deluxetable}

Comparing with other SNe IIn, e.g.\ SN 2008iy \citep{Miller_2008iy} which had a
measured X-ray luminosity of $L_X = \SI{2.4 \pm 0.8e41}{erg.\second^{-1}}$ or
SN 2010jl \citep{Ofek2014a} with $L_X \approx \SI{1.5e41}{erg.\second^{-1}}$, we
cannot exclude X-ray emission from PTF12csy with our measured upper limit.
However, \citet{Svirski} suggest that $L_X$ be about $10^{-4}$ of the bolometric
luminosity at the time of the shock breakout. With the estimated
bolometric luminosity from Section~\ref{sec:sed} around the time of the first
Swift observations, this implies $L_X \approx \SI{6.4e38}{erg.\second^{-1}}$,
well below our X-ray limits.

\section{LOW-ENERGY FOLLOW-UP DATA}
\label{sec:lowenergy}

\subsection{Optical and UV Observations of PTF12csy}
\label{sec:obs}

During the follow-up program of the neutrino alert, the first
observations were done on 2012 April 03, 05, 07 and 09 (MJD 56020 to
56026) by PTF with the Palomar Samuel Oschin 48-inch telescope (P48) \citep{PTF_overview}, which is
a wide-field Schmidt telescope.
The images (see
Fig.~\ref{fig:subtraction})
revealed a so far undiscovered supernova, named \emph{PTF12csy},
at a magnitude of $\sim 18.6$ in the Mould $R$-band.
More photometric observations were carried out, with the P48 and the
Palomar 60-inch (P60) telescopes \citep{PTF_overview} at the Palomar Observatory in
California, and the Faulkes Telescope North (FTN) \citep{FTN_overview}
at Haleakala on Maui, Hawaii.
Spectroscopy was taken as well, with the Gemini North Multi-Object Spectrograph
(GMOS) \citep{GeminiNorth_GMOS} on the \SI{8}{m} Gemini North telescope (Mauna
Kea, Hawaii) on 2012 April 17 (MJD 56034) and with the Low-Resolution Imaging
Spectrometer (LRIS) \citep{KeckI_LRIS} on the \SI{10}{m} Keck I telescope (Kamuela,
Hawaii) on 2013 February 09 (MJD 56332), enabling the identification of the
supernova as a Type IIn SN with narrow emission lines.
The spectra are available
from WISeREP\footnote{\url{http://wiserep.weizmann.ac.il}} \citep{Yaron2012}.

P48 data were extracted using an aperture photometry pipeline and are calibrated
with 21 close-by SDSS stars \citep{Eisenstein2011,Gunn2006,Ahn2014,Alam2015}. The faint
host galaxy was subtracted and the upper limits are at the \SI{5}{\ensuremath{\sigma}} level.
P48 magnitudes are in the PTF natural AB magnitude system, which is similar,
but not identical to the SDSS system. The difference is given by a color term,
which is ignored in this work, except for the conversion of Mould $R$ to SDSS
$r$, explained in Section~\ref{sec:corr}.
The P60 photometry is tied to the same 21 SDSS calibration stars. Note
that there might be host galaxy contamination in the late-epoch P60 photometry.
P60 magnitudes are in the SDSS AB magnitude system.
The FTN data were processed by an automatic pipeline, without host subtraction,
and agree very well with the host-subtracted P60 data taken in the same night.

The Pan-STARRS1 (PS1) telescope was not part of the real-time
triggering and response system, but its wide-field coverage provides a
useful archive to search retrospectively for detections. PS1 is a
$\SI{1.8}{m}$ telescope located at Haleakala on Maui in the Hawaiian
islands, equipped with a \SI{3.3}{\degree} FOV and a 1.4 gigapixel
camera \citep{Pan-STARRS}. In the course of its 3$\pi$ steradian
survey, the telescope observes each part of the sky typically \numrange{8}{10}
times per year \citep{magnier2013}.
PS1 first detected PTF12csy on MJD \num{55847.582} and archived it as object
\emph{PSO J104.6365+17.2622}.
The magnitudes in all PS1 images were obtained with PSF fitting
within the Pan-STARRS Image Processing Pipeline \citep{magnier2006}. They are
calibrated to typically seven local SDSS DR8 field stars. The magnitudes are in
the natural PS1 AB system as defined in \citet{tonry2012}, which is similar,
but not exactly the same as SDSS AB magnitudes. Particularly the $g$-band can
differ.

The Swift UVOT data were analyzed using the publicly available Swift analysis
tools \citep{HEAsoft}\footnote{See \url{http://www.swift.ac.uk/analysis/uvot/} for instructions},
and a source was seen close to the detection threshold.

ROTSE's limiting magnitude of about \SIrange{16}{17}{mag} prevented a
detection of the SN in ROTSE follow-up observations.


\subsection{Photometry}
\label{sec:phot}

\subsubsection{Photometric Corrections}
\label{sec:corr}

The photometry is corrected for Galactic extinction using $R_V = A_V / E(B-V) = 3.1$ and
$E(B-V) = 0.071$
\citep{Schlegel1998}\footnote{Obtained via
the NASA/IPAC Infrared Science Archive
\url{http://irsa.ipac.caltech.edu/applications/DUST/}.}. The extinction
coefficient is converted to the filters' effective wavelengths using the
algorithm from \citet{Cardelli1989}\footnote{Via
\url{http://dogwood.physics.mcmaster.ca/Acurve.html}.}, i.e.\ \SI{0.35}{mag}
for $u$, \SI{0.30}{mag} for $B$, \SI{0.26}{mag} for $g$, \SI{0.19}{mag} for
$r$, \SI{0.11}{mag} for $z$.
The extinction within the host galaxy could not be determined.

\begin{figure}[tbp]
    \plotone{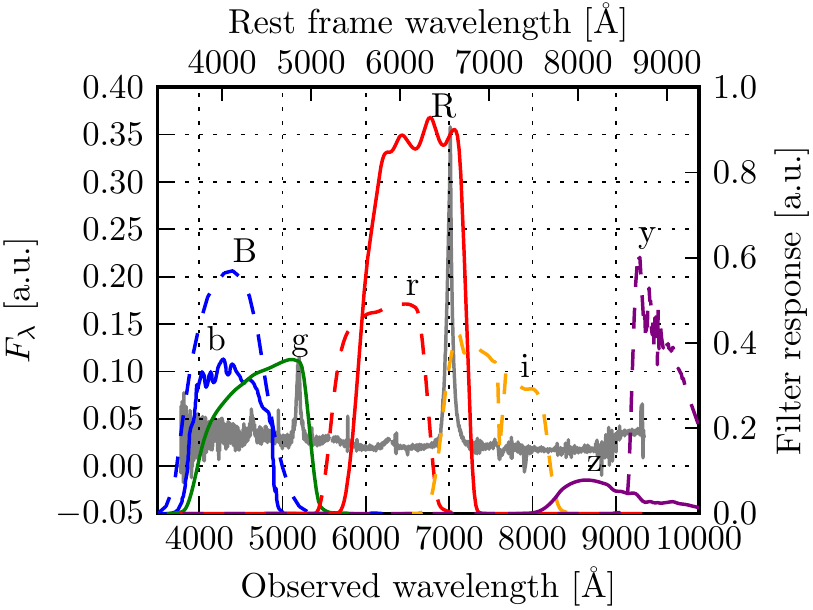}
    \caption{\emph{Background, gray, left axis:} Gemini North spectrum from
    2012 April 17 (MJD 56034). \emph{Foreground, multiple colors, right axis:}
    The filter response functions of the applied photometric filters, defined as
    filter transmission or effective area. Note that the absolute normalization
    is arbitrary and only the shape of the curves is relevant.}
    \label{fig:spectrum_and_filters}
\end{figure}

In Fig.~\ref{fig:spectrum_and_filters}, the Gemini North spectrum is
overlaid with the applied photometric filters. The strong Balmer lines
contribute differently to the various filters. For the SED
construction (see Section~\ref{sec:sed}), in order to approximate the
black body continuum, the contribution of the strongest emission
lines, H$\alpha$ and H$\beta$, is removed from the photometry using
the Gemini North spectrum and the filter curves. For
Figs.~\ref{fig:lc_abs} and~\ref{fig:decline},
the P48 Mould $R$ magnitudes are converted to SDSS $r$ by subtracting the
H$\alpha$ contribution (as above), applying the formulae in \citet{Ofek2012}
valid for black body spectra, and then re-adding the H$\alpha$
contribution to the $r$-band. After conversion, the P48 $R$ magnitudes are
consistent with the P60 SDSS $r$ magnitudes.

The Swift UVOT data contain host contamination. Since no GALEX data
from a pre- or post-SN epoch are available for the host
galaxy\footnote{See \url{http://galex.stsci.edu/GalexView/}},
no host subtraction can be done in the UV filters of UVOT\@. For the $u$, $b$ and
$v$ filters, the host is subtracted by interpolating the host magnitudes from
the SDSS DR12
data \citep{Alam2015}\footnote{\url{http://skyserver.sdss.org/dr12/en/tools/chart/navi.aspx}}
to the effective wavelengths of the UVOT filters.

\subsubsection{The Light Curves}
\label{sec:lc}

The earliest detection of PTF12csy was in the Pan-STARRS1 y-band on
2011 October 13 (MJD \num{55847.582}), 169 days prior to the neutrino alert in
observer frame, corresponding to 158 days in host galaxy rest frame using
$z=0.0684$ (see Section~\ref{sec:spec}). The latest non-detection,
again in Pan-STARRS1, was on 2011 March 21 (MJD 55641.3) in a
$\SI{30}{s}$ $z$-band frame, 206 days before the first detection
(193 days in rest frame). Hence, the explosion time is not well
constrained and can be anytime between MJD 55641.3 and MJD \num{55847.6}.
Hereafter, we refer to the $y$-band detection at MJD \num{55847.582} as the
first detection and use it as day 0 for the light curve.

\begin{figure}[tbp]
    \plotone{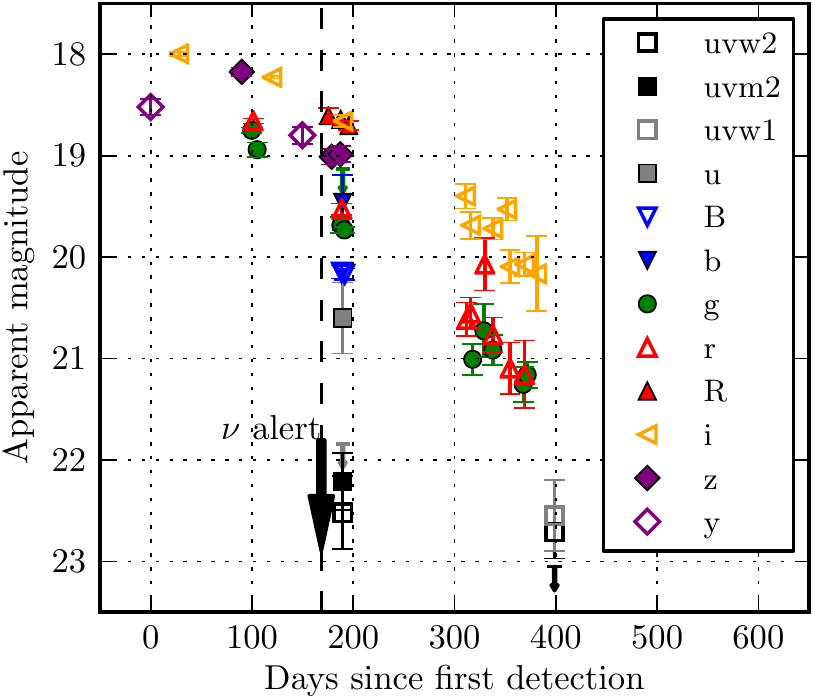}
    \caption{PTF12csy photometry in apparent magnitudes without applying
    corrections. The photometry is averaged over intervals of 10
    days. The data originate from the following telescopes: $uvw2$, $uvm2$,
    $uvw1$, $u$, $b$: UVOT; $B$: P60; $g$: P60, PS1, FTN; $r$: P60, PS1, FTN; $R$: P48;
    $i$: P60, PS1, FTN; $z$: P60, PS1; $y$: PS1.}
    \label{fig:lc_app}
\end{figure}

The uncorrected SN light curves with the data available through the IceCube
optical follow-up program are displayed in Fig.~\ref{fig:lc_app}, including
photometry acquired with the Swift UVOT filters $uvw2$, $uvm2$,
$uvw1$, $u$ and $b$; the Johnson $B$ filter on P60; the SDSS filters
$g$, $r$, $i$ with data from P60, PS1, and FTN; the SDSS $z$ filter on
P60; Mould $R$ filter on P48; and Pan-STARRS $y$ filter on PS1.
The entire uncorrected photometry in apparent magnitudes, as seen in
Fig.~\ref{fig:lc_app}, is also available in Table~\ref{tab:lc}. The
light curves are averaged within bins of 10 days width, for each
filter and telescope separately.
Note that, in contrast to most of the other photometry, no host
subtraction is performed for the Swift UVOT magnitudes presented
in Fig.~\ref{fig:lc_app} and Table~\ref{tab:lc}.

\begin{deluxetable*}{ccccccc}
    \tablewidth{\textwidth}
    \tablecaption{Photometric observations of PTF12csy\label{tab:lc}}
    \tablehead{\colhead{MJD Date} & \colhead{Rest frame days} & \colhead{Mag} &
        \colhead{Abs.\ mag} & \colhead{Lim.\ mag} & \colhead{Filter} & \colhead{Tel.}}
    \startdata
        \num{55273.219} & \num{-537.589} & \nodata & \nodata & \num{21.11} & $g$ & P48 \\
        \num{55294.162} & \num{-517.987} & \nodata & \nodata & \num{20.62} & $R$ & P48 \\
        \num{55431.514} & \num{-389.429} & \nodata & \nodata & \num{18.99} & $R$ & P48 \\
        \num{55477.402} & \num{-346.478} & \nodata & \nodata & \num{20.67} & $R$ & P48 \\
        \num{55596.889} & \num{-234.642} & \nodata & \nodata & \num{18.90} & $g$ & P48 \\
        \num{55641.304} & \num{-193.071} & \nodata & \nodata & \num{21.40} & $z$ & PS1 \\
        \num{55847.588} & \num{0.005} & \num{18.52 \pm 0.08} & \num{-18.92} & \nodata & $y$ & PS1 \\
        \num{55875.515} & \num{26.145} & \num{18.00 \pm 0.02} & \num{-19.45} & \nodata & $i$ & PS1 \\
        \num{55937.502} & \num{84.163} & \num{18.17 \pm 0.04} & \num{-19.27} & \nodata & $z$ & PS1 \\
        \num{55948.816} & \num{94.752} & \num{18.82 \pm 0.05} & \num{-18.62} & \nodata & $g$ & PS1 \\
        \num{55948.841} & \num{94.776} & \num{18.66 \pm 0.02} & \num{-18.79} & \nodata & $r$ & PS1 \\
        \num{55957.475} & \num{102.858} & \num{19.00 \pm 0.02} & \num{-18.44} & \nodata & $g$ & PS1 \\
        \num{55967.269} & \num{112.024} & \num{18.23 \pm 0.01} & \num{-19.22} & \nodata & $i$ & PS1 \\
        \num{55997.366} & \num{140.194} & \num{18.80 \pm 0.08} & \num{-18.64} & \nodata & $y$ & PS1 \\
        \num{56022.981} & \num{164.169} & \num{18.61 \pm 0.08} & \num{-18.84} & \num{18.23} & $R$ & P48 \\
        \num{56026.246} & \num{167.225} & \num{19.01 \pm 0.08} & \num{-18.43} & \nodata & $z$ & PS1 \\
        \num{56034.841} & \num{175.270} & \num{19.47 \pm 0.04} & \num{-17.98} & \nodata & $r$ & P60 \\
        \num{56034.844} & \num{175.273} & \num{18.99 \pm 0.08} & \num{-18.46} & \nodata & $z$ & P60 \\
        \num{56035.177} & \num{175.584} & \num{18.64 \pm 0.04} & \num{-18.80} & \nodata & $R$ & P48 \\
        \num{56035.922} & \num{176.281} & \num{18.68 \pm 0.03} & \num{-18.76} & \nodata & $i$ & P60 \\
        \num{56035.925} & \num{176.284} & \num{20.15 \pm 0.06} & \num{-17.30} & \nodata & $B$ & P60 \\
        \num{56035.928} & \num{176.287} & \num{19.69 \pm 0.05} & \num{-17.75} & \nodata & $g$ & P60 \\
        \num{56036.581} & \num{176.899} & \num{18.66 \pm 0.04} & \num{-18.78} & \num{20.97} & $i$ & FTN \\
        \num{56036.585} & \num{176.902} & \num{19.54 \pm 0.06} & \num{-17.90} & \num{21.12} & $r$ & FTN \\
        \num{56036.590} & \num{176.906} & \num{19.73 \pm 0.07} & \num{-17.72} & \num{21.23} & $g$ & FTN \\
        \num{56037.150} & \num{177.431} & \nodata & \nodata & \num{19.13} & $v$ & UVOT \\
        \num{56037.150} & \num{177.431} & \num{20.60 \pm 0.35} & \num{-16.84} & \num{20.79} & $u$ & UVOT \\
        \num{56037.150} & \num{177.431} & \num{22.21 \pm 0.28} & \num{-15.23} & \num{22.76} & $uvm2$ & UVOT \\
        \num{56037.150} & \num{177.431} & \num{22.52 \pm 0.36} & \num{-14.92} & \num{22.71} & $uvw2$ & UVOT \\
        \num{56037.150} & \num{177.431} & \nodata & \nodata & \num{21.84} & $uvw1$ & UVOT \\
        \num{56037.150} & \num{177.431} & \num{19.46 \pm 0.27} & \num{-17.98} & \num{19.95} & $b$ & UVOT \\
        \num{56039.175} & \num{179.326} & \num{18.74 \pm 0.03} & \num{-18.70} & \nodata & $i$ & P60 \\
        \num{56039.177} & \num{179.328} & \num{19.55 \pm 0.04} & \num{-17.89} & \nodata & $r$ & P60 \\
        \num{56039.178} & \num{179.329} & \num{20.17 \pm 0.06} & \num{-17.28} & \nodata & $B$ & P60 \\
        \num{56039.181} & \num{179.332} & \num{19.73 \pm 0.04} & \num{-17.71} & \nodata & $g$ & P60 \\
        \num{56040.285} & \num{180.365} & \num{18.60 \pm 0.07} & \num{-18.85} & \num{20.98} & $i$ & FTN \\
        \num{56043.178} & \num{183.073} & \num{18.70 \pm 0.05} & \num{-18.74} & \nodata & $R$ & P48 \\
        \num{56158.504} & \num{291.015} & \num{19.40 \pm 0.12} & \num{-18.05} & \nodata & $i$ & P60 \\
        \num{56161.496} & \num{293.815} & \num{20.58 \pm 0.12} & \num{-16.86} & \nodata & $r$ & P60 \\
        \num{56163.490} & \num{295.682} & \num{19.69 \pm 0.13} & \num{-17.76} & \nodata & $i$ & P60 \\
        \num{56165.486} & \num{297.550} & \num{21.01 \pm 0.15} & \num{-16.44} & \nodata & $g$ & P60 \\
        \num{56176.465} & \num{307.826} & \num{20.73 \pm 0.26} & \num{-16.72} & \nodata & $g$ & P60 \\
        \num{56177.453} & \num{308.751} & \num{20.07 \pm 0.26} & \num{-17.37} & \nodata & $r$ & P60 \\
        \num{56185.431} & \num{316.218} & \num{19.72 \pm 0.11} & \num{-17.73} & \nodata & $i$ & P60 \\
        \num{56185.432} & \num{316.219} & \num{20.77 \pm 0.17} & \num{-16.68} & \nodata & $r$ & P60 \\
        \num{56185.436} & \num{316.223} & \num{20.92 \pm 0.15} & \num{-16.53} & \nodata & $g$ & P60 \\
        \num{56200.887} & \num{330.685} & \num{19.81 \pm 0.30} & \num{-17.63} & \nodata & $i$ & P60 \\
        \num{56202.385} & \num{332.086} & \num{21.09 \pm 0.25} & \num{-16.35} & \nodata & $r$ & P60 \\
        \num{56215.381} & \num{344.250} & \num{21.23 \pm 0.39} & \num{-16.21} & \nodata & $r$ & P60 \\
        \num{56215.429} & \num{344.295} & \num{20.11 \pm 0.10} & \num{-17.33} & \nodata & $i$ & P60 \\
        \num{56215.435} & \num{344.301} & \num{21.25 \pm 0.17} & \num{-16.19} & \nodata & $g$ & P60 \\
        \num{56219.344} & \num{347.960} & \num{21.16 \pm 0.13} & \num{-16.28} & \nodata & $g$ & P60 \\
        \num{56219.837} & \num{348.421} & \num{20.97 \pm 0.13} & \num{-16.47} & \nodata & $r$ & P60 \\
        \num{56224.988} & \num{353.242} & \num{20.01 \pm 0.17} & \num{-17.44} & \nodata & $i$ & P60 \\
        \num{56229.808} & \num{357.754} & \num{20.20 \pm 0.45} & \num{-17.24} & \nodata & $i$ & P60 \\
        \num{56246.320} & \num{373.208} & \num{22.71 \pm 0.26} & \num{-14.73} & \num{23.34} & $uvw2$ & UVOT \\
        \num{56246.320} & \num{373.208} & \num{22.55 \pm 0.35} & \num{-14.89} & \num{22.79} & $uvw1$ & UVOT \\
        \num{56246.320} & \num{373.208} & \nodata & \nodata & \num{23.05} & $uvm2$ & UVOT \\
    \enddata
    \tablecomments{Rest frame days relative to first detection on MJD
    \num{55847.582}. 10-day binning. No correction for extinction.}
\end{deluxetable*}

\begin{figure}[tbp]
    \plotone{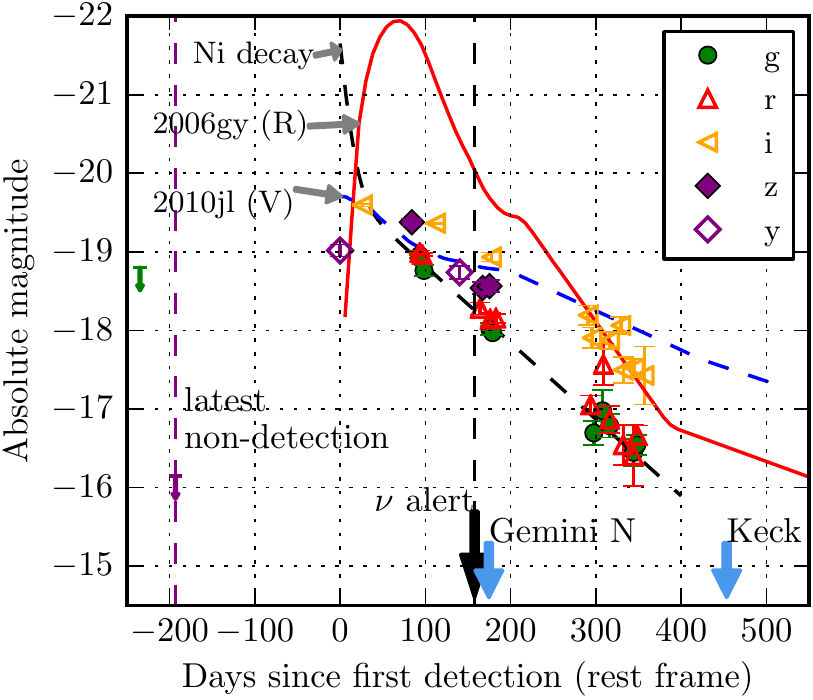}
    \caption{PTF12csy photometry (symbols) in absolute magnitudes,
    with correction for Galactic extinction, and conversion of P48 Mould R
    magnitudes to SDSS $r$ magnitudes (see Section~\ref{sec:corr}).
    The data originate from the following telescopes:
    $g$: P48, P60, PS1, FTN; $r$: P48, P60, PS1, FTN; $i$: P60, PS1, FTN; $z$: P60, PS1; $y$: PS1.
    The photometry is averaged over intervals of 10 days.
    Other absolute SN II light curves (lines) and a theoretical light curve from
    radioactive decay of nickel (black dashed line) are added for comparison.
    The comparison light curves are partly not extinction corrected and have
    different reference dates (see text).
    }
    \label{fig:lc_abs}
\end{figure}

Figure~\ref{fig:lc_abs} shows the light curve of selected filters in absolute magnitudes,
after the photometric corrections. Light curves of other exceptional
Type II SNe are overlaid for comparison: SN IIn 2006gy \citep{Ofek_2006gy,
Smith_2006gy, Kawabata_2006gy, Miller_2006gy}, one of the most luminous SNe
ever recorded, and SN 2010jl \citep{Stoll_2010jl, Zhang_2010jl}, a SN IIn
that is spectroscopically similar to PTF12csy (see Section~\ref{sec:spec})
and shows signs of a collisionless shock in an optically thick circumstellar medium
(CSM), hinting towards potential high-energy neutrino production \citep{Ofek2013}.
Note that the
SN 2010jl light curve is not extinction corrected and the comparison light curves
have different reference dates: SN 2010jl is relative to maximum light, 2006gy
is relative to the explosion time.
A theoretical light curve from pure radioactive decay of ${}^{56}\text{Ni}
\rightarrow {}^{56}\text{Co} \rightarrow {}^{56}\text{Fe}$ (black dashed line)
is added to the figure as well, scaled to match the observed absolute magnitude
of PTF12csy.

The brightest observed absolute magnitudes after application of photometric
corrections (see Section~\ref{sec:corr}) and converting to absolute magnitudes
with a distance modulus of $\mu=37.443$ ($z=0.0684$) are $M_g \approx
\SI{-19.0}{mag}$, $M_r \approx \SI{-19.0}{mag}$, $M_i \approx \SI{-19.6}{mag}$,
$M_z \approx \SI{-19.4}{mag}$, and $M_y \approx \SI{-19.0}{mag}$, assuming
standard cosmology with Hubble parameter $H_0 =
\SI{70}{km.\second^{-1}.Mpc^{-1}}$, matter density $\Omega_m = 0.3$, and dark
energy density $\Omega_\Lambda = 0.7$. While these are lower limits to the peak
magnitude due to the sparse sampling, these absolute magnitudes are relatively
modest compared to the most luminous SNe IIn, e.g.\ SN 2006gy ($M_R =
\SI{-22}{mag}$) \citep{Kawabata_2006gy} or SN 2008fz ($M_V = \SI{-22.3}{mag}$)
\citep{Drake_2008fz}. They are however comparable to the SNe IIn 2008iy ($M_r
\approx \SI{-19.1}{mag}$) \citep{Miller_2008iy}, 1988Z ($M_R \lesssim
\SI{-18.9}{mag}$) \citep{Turatto_1988Z} and SN 2010jl ($M_R \lesssim
\SI{-20.0}{mag}$) \citep{Zhang_2010jl}.

\subsubsection{Decline Rates and Energy Source}
\label{sec:decline}

\begin{figure}[tbp]
    \plotone{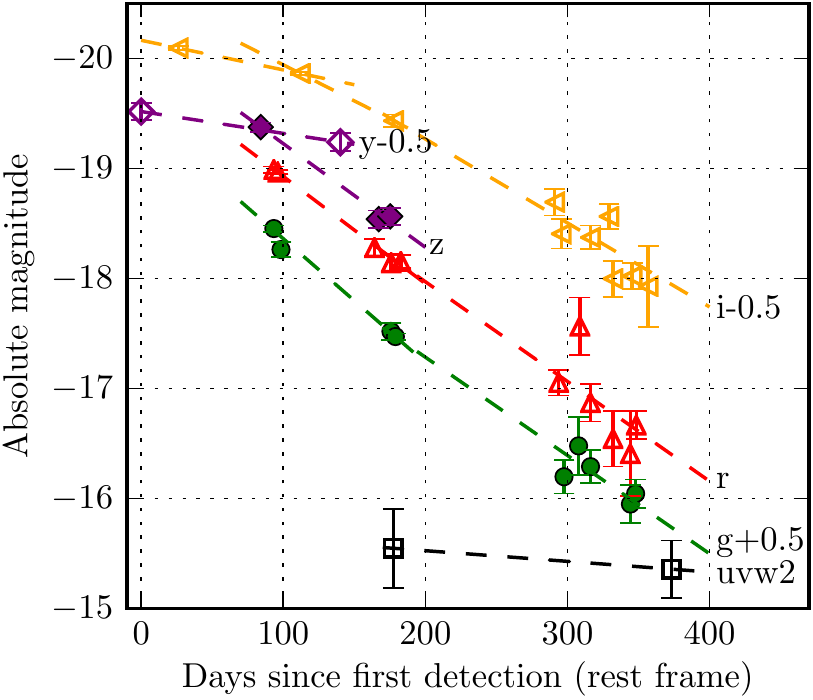}
    \caption{Light curves of several filters with the fitted linear
    declines. See Table~\ref{tab:decline} for the numerical values of the
    found decline rates.}
    \label{fig:decline}
\end{figure}

The light curves of PTF12csy indicate a plateau within $\sim
\SI{100}{days}$ after first detection, and a slow fading afterwards.
The corrected absolute magnitude light curves are fitted to
obtain the linear decline rates in different photometric filters, during
different epochs (see Fig.~\ref{fig:decline} and Table~\ref{tab:decline}). For
some epochs and filters, especially $g$ and $r$, the decline rates are close to
\SI{0.98}{mag.(100 days)^{-1}}, the decline rate expected for radioactive
$^{56}$Co decay \citep{Miller_2008iy}, while in general decline rates are
slower, indicating that at least part of the radiated energy is powered by
interaction of the SN ejecta with a dense CSM \citep{Miller_2008iy}.

\begin{deluxetable}{cccc}
    \tablecaption{Decline rates of the PTF12csy light curve\label{tab:decline}}
    \tablehead{\colhead{Filter} & \colhead{\SIrange{0}{150}{d}\tablenotemark{a}} &
        \colhead{\SIrange{70}{200}{d}\tablenotemark{a}} & \colhead{\SIrange{170}{400}{d}\tablenotemark{a}}}
    \startdata
        $uvw2$ & \nodata & \nodata & \num{0.097\pm0.227} \\
        $g$ & \nodata & \num{1.127\pm0.044} & \num{0.893\pm0.051} \\
        $r$ & \nodata & \num{0.974\pm0.053} & \num{0.907\pm0.059} \\
        $i$ & \num{0.269\pm0.024} & \num{0.656\pm0.089} & \num{0.764\pm0.052} \\
        $z$ & \nodata & \num{0.943\pm0.079} & \nodata \\
        $y$ & \num{0.199\pm0.080} & \nodata & \nodata
    \enddata
    \tablenotetext{a}{Units: $\si{mag.(\SI{100}{d})^{-1}}$. Indicated periods in
    rest frame days relative to first detection on MJD \num{55847.582}.}
\end{deluxetable}

Additionally, radioactive decay of $^{56}$Co at a still relatively
high absolute magnitude of $\sim \SI{-19}{mag}$ implies a preceding
$^{56}$Ni decay with an extremely bright peak, which was not observed,
although the data are quite sparse.
Assuming that the luminosity is generated by radioactive decay alone and
following \citet{Kulkarni2005}, we estimate that $\gtrsim \SI{1.7}{\solarmass}$
of $^{56}$Ni would be required to provide the bolometric luminosity of
\SI{9.7e42}{erg.\second^{-1}} at \SI{100}{d} in rest frame (see
Section~\ref{sec:sed}). The lower limit on the $^{56}$Ni mass is set by assuming
that the explosion and thus generation of $^{56}$Ni was at the latest possible
time, directly before the first detection.
Figure~\ref{fig:lc_abs} shows the corresponding theoretical light curve
resulting from the radioactive decay of nickel and cobalt (black dashed line).
Adopting an earlier explosion time results in an even larger $^{56}$Ni mass.

This is much more than the usual amount of
$^{56}$Ni of \SI{<0.5}{\solarmass}, often \SI{<0.1}{\solarmass} (see e.g.\
\citet{NickelMasses}, also \citet{Margutti_2009ip}). However, extremely
superluminous SNe might have $^{56}$Ni masses of that order of magnitude
\citep{GalYam_Luminous}.

We note that in addition to the $^{56}$Ni mass and luminosity arguments, the
spectrum showing intermediate width Balmer lines and a continuum appears
inconsistent with radioactive decay as well (see Section~\ref{sec:spec}).

\subsubsection{Fitting to an Interaction Model}
\label{sec:fit}

Here we assume that the light curve is powered by conversion of the ejecta's
kinetic energy to luminosity through interaction of the ejecta with the CSM\@.
Following \citet[see also: e.g. \citet{Chugai1994, Svirski,
Moriya2013}]{Ofek2014a}, we model the light curve as a power-law of the form
$L(t) = L_0 t^\alpha$.

After shock breakout, there is a phase of power-law decline of the
luminosity, with an index of typically $\alpha \approx -0.3$.
This lasts until the shock runs over a CSM mass equivalent to the
ejecta mass and the shock enters a new phase of either conservation of
energy if the density is low enough and the gas cannot cool quickly
(the Sedov-Taylor phase), or conservation of momentum if the gas
radiates its energy via fast cooling (the snow-plow phase). During the
late stage, the light curve will be declining more steeply, in both
cases \citep{Ofek2014a}.

Since we assume PTF12csy to be powered by interaction, we try to fit
the interaction model from \citet{Ofek2014a} to the light curve data
with the least-squares method. We perform the fit within the range
of \numrange{93}{200} rest frame days, starting at the first r-band
detection, and use the r-band light curve scaled with the bolometric
luminosity from Section~\ref{sec:sed}. It is found that the
power-law index $\alpha$ needs to be significantly steeper than
\num{-0.3} in order to reasonably describe the data. It lies in the
range of \numrange{-3}{-1.2}, using the constraint on the explosion
time (see Section~\ref{sec:lc}) for the temporal zero point of the
power-law. The best fit is at $\alpha=-3$, with the explosion time at
the lowest allowed value which is the date of the last non-detection.
This suggests a very steep CSM density profile $\propto r^{-5}$
\citep[see][eq.~12]{Ofek2014a}, compared to the profile $\propto
r^{-2}$ resulting from a wind with steady mass loss.
However, the
self-similar solutions of the hydrodynamical equations \citep{Chevalier1982}
that are used in \citet[eq.~12]{Ofek2014a} are invalid if the CSM density
profile is steeper than $r^{-3}$. But nevertheless, as discussed for the
late-time light curve in \citet[sec.~5.2]{Ofek2014a}, probably the profile
is steeper than $r^{-3}$.

This leaves us with several possible explanations:
\begin{enumerate}
    \item Already between rest frame days 93 and 200, the SN was in
        the late, e.g.\ snow-plow, phase.
        This is consistent with SN 2010jl, where the late-time light curve
        also shows a power-law index $\alpha \approx -3$ \citep[sec.~5.2]{Ofek2014a}.
        Assuming that the break in
        the light curve between power-law phase and late phase
        occurred just before the first r-band detection at day 93, and
        comparing with SN 2010jl \citep{Ofek2014a}, then this means that the
        SN was likely already a few hundred days old, and the r-band
        maximum was about \numrange{1}{1.25} \si{mag} brighter than
        the observed one \citep[cf.][Fig.~1]{Ofek2014a},
        consistent with SN 2010jl's r-band maximum. It follows that
        the power-law phase ended $\leq 286$ rest frame days after
        explosion, from which we can derive a swept CSM mass of
        $\lesssim \SI{12}{\solarmass}$, using \citet[eq.~22]{Ofek2014a} and
        adopting the standard values
        given for SN 2010jl.
    \item The SN is powered by ejecta-CSM interaction, but its light
        curve is declining steeper than a $t^{-0.3}$ power-law. This
        is possible, e.g.\ if spherical symmetry, assumed in
        \citet{Ofek2014a}, is broken,
        if the optical depth is lower than in SN 2010jl
        or if the CSM density profile falls steeper than $r^{-2}$
        (s.a.).
    \item The SN is not powered by interaction, but by radioactive
        decay, leading to an exponential light curve decline. However,
        this appears unlikely, as noted above in
        Section~\ref{sec:decline}.
\end{enumerate}

\subsubsection{Spectral Energy Distribution (SED)}
\label{sec:sed}

\begin{figure}[tbp]
    \plotone{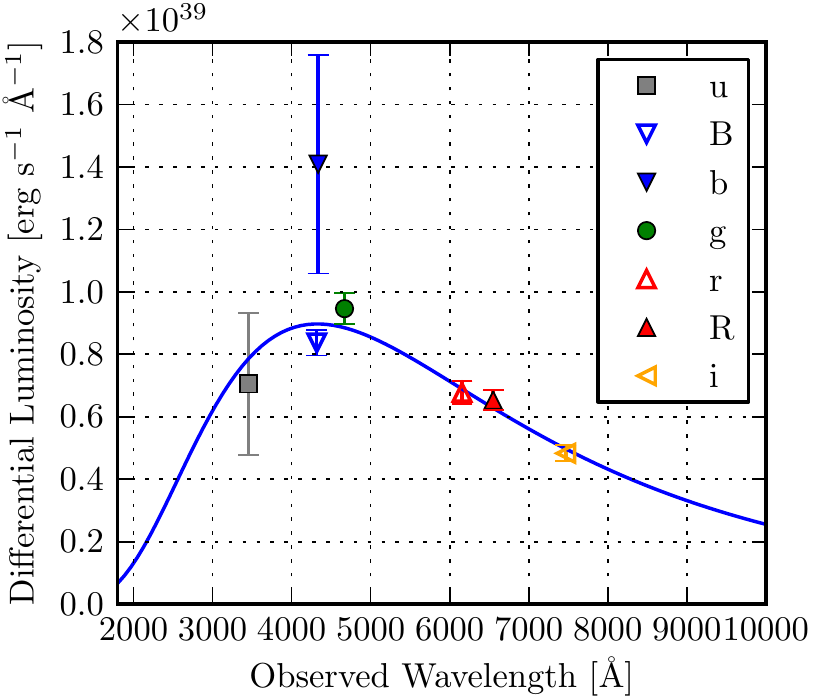}
    \caption{SED of PTF12csy using photometry from 10 days around
    day 189 (observer frame) after the first detection.
    The fitted rest frame temperature is $T=\SI{7160 \pm 270}{K}$ and
    the fitted bolometric luminosity \SI{5.53 \pm 1.18e+42}{erg.\second^{-1}}.
    }
    \label{fig:sed}
\end{figure}

Since the spectra are only roughly calibrated, the spectral energy distribution
(SED) is approximated from photometric data. For the highest spectral
range and number of observations, a window of 10 observer frame days
around day 189, from day 184 to 194, is used to select data (day
172.2 to day 181.6 in rest frame). Photometric corrections are
applied, e.g. H$\alpha$ and H$\beta$ removed (see
Section~\ref{sec:corr}). The data are plotted in
Fig.~\ref{fig:sed} as function of the filters' effective
wavelengths. A black body spectrum is assumed to describe the SED and
is fitted to the data. For each filter, a
model data point corresponding to the black body spectrum is calculated
via integration of the black body spectrum and the
filter function following the SDSS definition of AB magnitude in
\citet[eq.~7]{SDSS_ABmag}. A $\chi^2$ fit minimizes the difference between the
model data and the measured data.

The fit results in a reduced $\chi^2 / n_\text{dof} = \num{7.9} / 5 = \num{1.6}$ and delivers
estimates for both the rest frame temperature $T$ and the absolute
bolometric luminosity $L_\text{bol}$ of the photosphere emitting the black body
radiation: $T = \SI{7160 \pm 270}{K}$ and $L_\text{bol} = \SI{5.53 \pm
1.18e+42}{erg.\second^{-1}}$,
where the errors correspond to \SI{1}{\ensuremath{\sigma}}.
Applying the Stefan-Boltzmann law, we can calculate the radius of the
black body photosphere from the bolometric luminosity. We estimate it
to be $R_\text{phot} = \SI{1.7 \pm 0.1e+15}{cm}$.

Finally, to obtain an estimate on the total radiated energy, the
lines' contributions to the luminosity have to be added to the
continuum luminosity. Using the Gemini North spectrum, the
contribution of the H$\alpha$ and H$\beta$ line to the total
luminosity is computed and added to the continuum luminosity from the
black body fit. This results in an estimated total radiated luminosity
of \SI{6.4 \pm 1.2e+42}{erg.\second^{-1}} at day 189 in the observer frame, i.e.\ day
177 in the rest frame.

We use the fitted shape of the $i$-band light curve
(cf.~Section~\ref{sec:phot}, Fig.~\ref{fig:decline},
Table~\ref{tab:decline}) to extrapolate this value and find a total
radiated luminosity of \SI{\sim 9.7e+42}{erg.\second^{-1}} at \SI{100}{d}
(rest frame), as used in Section~\ref{sec:decline},
and a total energy of $E_\mathrm{bol} = \SI{2.1e+50}{erg}$ radiated within
400 rest frame days after first detection, comparable to SN
2008iy which had \SI{\sim 2e50}{erg}) \citep{Miller_2008iy} and SN 2010jl
with \SI{4.3e50}{erg} \citep{Zhang_2010jl}.
This is a lower limit on the total radiated energy, since we lack photometric
data between explosion and first detection and do not extrapolate before the
first detection. Additionally, as discussed below, we are neglecting a
possible contribution of X-ray and $\gamma$-ray emission to the total radiated
energy, which is not considered by the black body spectrum based on the UV and
optical data.

We recommend to treat these results with caution, since
\citet{Ofek2014a} pointed out that at late times the fraction of energy
released from SNe IIn in X-rays can increase, causing the optical
spectrum to deviate from a black body as fewer photons are available
in the optical. This can lead to an effective decrease of the
estimated photospheric radius.
In this context, our estimates of $R_\mathrm{phot}$, $L_\mathrm{bol}$,
and $E_\mathrm{bol}$
must be treated as lower limits.
Unfortunately, the X-ray flux from PTF12csy was not detected (see
Section~\ref{sec:x-ray}).

\subsection{Spectroscopy}
\label{sec:spec}

\begin{figure*}
    \begin{center}
        \includegraphics[width=\textwidth]{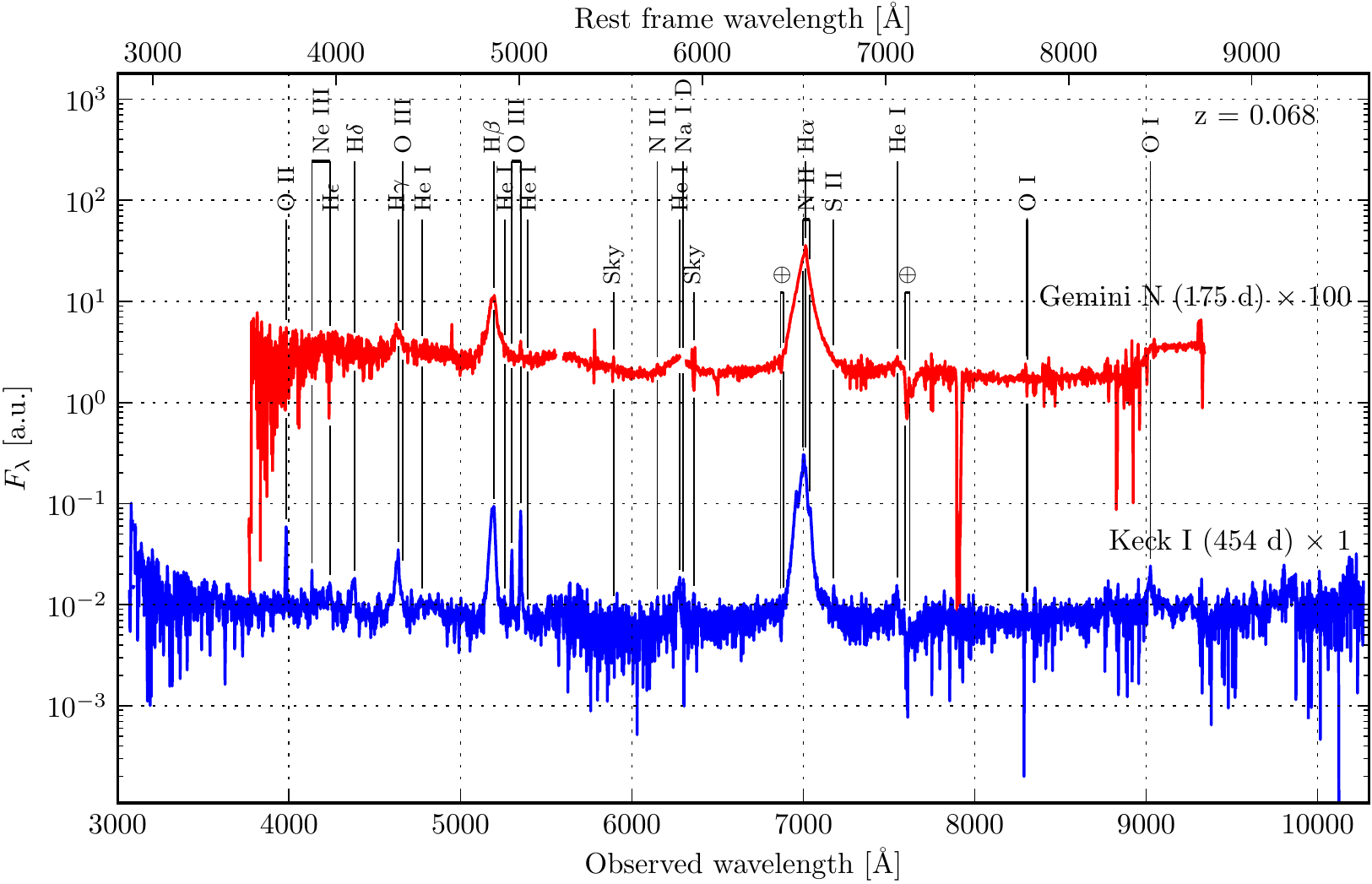}
    \end{center}
    \caption{Spectra taken with Gemini North on 2012 April 17 (top) and
    Keck I on 2013 February 09, showing narrow (Type IIn) emission lines.
    The H$\alpha$ line at $\sim \SI{7000}{\angstrom}$ (observer
    frame) is the strongest emission line and has a complicated
    structure. See Fig.~\ref{fig:halpha} for a close-up of the
    H$\alpha$ line.
    }
    \label{fig:spectra}
\end{figure*}

Two spectra were acquired (Table~\ref{tab:speclog}, Fig.~\ref{fig:spectra}).
They are dominated by narrow
emission lines, characteristic for Type IIn SNe, with a very weak blue
continuum, which indicates the old age of the SN\@. No continuum is
visible in the late spectrum. The SN emission lines are primarily
hydrogen, the Balmer series is visible from H$\alpha$ up to
H$\epsilon$. The oxygen lines \species{O}{i} $\lambda\lambda
7772,7774,7775,8447$, \species{O}{ii} $\lambda 3727$ with FWHM $\approx
\SI{500}{km.\second^{-1}}$, and \species{O}{iii} $\lambda\lambda 4364,4960,5008$ with
FWHM $\approx \SI{350}{km.\second^{-1}}$ are very narrow and were most
likely produced by circumstellar gas released by the progenitor prior
to explosion and then photoionized by UV radiation
\citep{Filippenko1997}.

\begin{deluxetable}{ccccc}
    \tablewidth{\columnwidth}
    \tablecaption{Log of spectral observations\label{tab:speclog}}
    \tablehead{\colhead{MJD} & \colhead{$T_\text{disc}$} &
        \colhead{$T_\text{det}$} &
        \colhead{\parbox{1.4cm}{\centering $\Delta v$ (\si{km.s^{-1}})}} &
        \colhead{Instrument}}
    \startdata
        56034   & 11              & 175            & 80               & Gemini North GMOS \\
        56332   & 290             & 454            & 100              & Keck I LRIS
    \enddata
    \tablecomments{$T_\text{disc}$: Rest frame days after PTF discovery. $T_\text{det}$:
    Rest frame days after first detection by PS1. $\Delta v$: spectral resolution at
    the H$\alpha$ line at \SI{7014}{\angstrom} in observer frame.}
\end{deluxetable}

\begin{figure}[tbp]
    \plotone{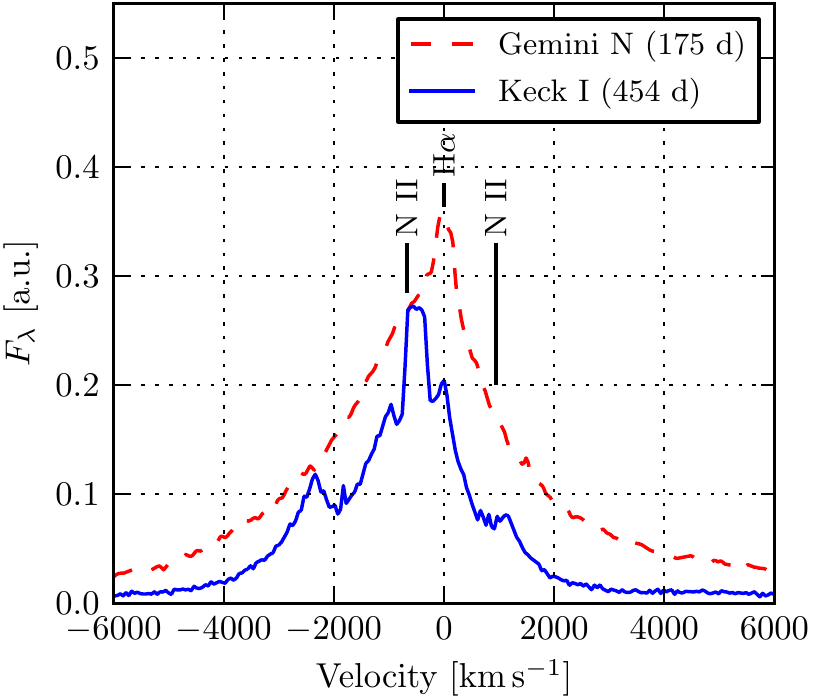}
    \caption{Comparison of the H$\alpha$ line in both spectra. The
    x-axis shows the Doppler velocity relative to the line center at
    $\SI{6564.61}{\angstrom}$, assuming a redshift of 0.0684.
    }
    \label{fig:halpha}
\end{figure}

Figure~\ref{fig:halpha} shows a close-up on the H$\alpha$ line from
both spectra, plotted vs.\ Doppler velocity relative to the rest frame
line center. In the early spectrum, the H$\alpha$ line peaks at the
line center and the line is composed of a narrow, intermediate and
broad component
with FWHM of $\sim \SI{400}{km.\second^{-1}}$,
$\sim \SI{2000}{km.\second^{-1}}$, and $\sim
\SI{5000}{km.\second^{-1}}$ respectively, found by fitting a
superposition of three Gaussian functions to the H$\alpha$ profile.
This is similar to other SNe IIn, e.g. SN 1988Z and 2008iy
\citep{Turatto_1988Z, Filippenko1997, Miller_2008iy}.
The H$\alpha$ profile with broad and intermediate-width component can be
explained as a result of the interaction of the SN ejecta
with a two-component wind \citep{Chugai1994}. In this model, the broad
line component is emitted from shocked SN ejecta expanding in a relatively
rarefied wind, while the intermediate component arises from a shocked
dense part of the wind, which can either consist of dense clumps or be
a dense equatorial wind \citep{Chugai1994}. This is an indication
of ejecta-CSM interaction.

The intermediate and broad component of the early spectrum's H$\alpha$ line are
blueshifted which may indicate formation of dust, as explored by
\citet{Smith2012} for SN 2010jl.
A more recent study by \citet{Gall2014}, the most comprehensive work on the SN
2010jl emission line blueshifts to date, finds very strong evidence for a
wavelength dependence of the blueshift. Therefore, the authors conclude that the
origin of the blueshifts is most likely the rapid formation of large dust
grains, confirming \citet{Smith2012}
and having implications on the origin of dust in galaxies.

Alternatively, \citet{Fransson2014} explain the line blueshift in SN 2010jl
with a bulk velocity of the emitting gas towards the observer.  This is more
consistent with observations if the spectral lines are symmetric about a center
and if there is no wavelength dependence of the blueshift.  The bulk velocity
is believed to be the result of radiative acceleration of the gas by flux from
the SN\@. Presumably, there are also other possible explanations for the
blueshift, e.g.\ the geometry or density structure of the CSM\@.
In case of SN PTF12csy, spectral line blueshift is only clearly visible in the
H$\alpha$ line, prohibiting the interpretation of the blueshift in favor of any
scenario.

The late spectrum's H$\alpha$ line has a much more complicated structure than
the early spectrum's. It does not peak at zero velocity anymore, but the peak is
blueshifted and there are many sub-peaks.
Again, the blueshift might be connected to
dust formation or radiative gas acceleration,
but other reasons are conceivable as well.
At least part of the late H$\alpha$ line's complex appearance might
be due to superposition of other spectral lines, e.g.\ \species{N}{ii}\@. Apart
from that, it is an indication of an inhomogeneous, maybe clumpy, CSM structure,
and perhaps asymmetric SN explosion.

The spectra are compared to template spectra from the
Padova-Asiago Supernova Archive (ASA) \citep{Harutyunyan2008} using the online tool
GELATO\footnote{\url{https://gelato.tng.iac.es}}.
The algorithm \citep{Harutyunyan2008} divides a spectrum into 11
relevant bins and averages within the bins to classify and compare
with the archived spectra. The PTF12csy spectra are de-reddened
with $E(B-V) = 0.1$ and compared to other Type II SNe. GELATO returns
the best 30 matching spectra together with their phases, ordered by
quality of fit.
For both the Gemini North spectrum taken at \SI{175}{d} and the Keck I spectrum
from \SI{454}{d}, the majority of best matching template spectra come from SN
2010jl.  The mean phase of the matching spectra is significantly higher for the
Keck I spectrum: \SI{378 \pm 102}{d}, versus \SI{154 \pm 21}{d} for the Gemini
N spectrum. However, the reference dates for the spectra are mostly the
discovery date, only rarely the date of maximum light or explosion date.

\subsection{Host Galaxy}
\label{sec:host}

The galaxy hosting PTF12csy is a faint dwarf galaxy designated SDSS
J065832.82+171541.6\footnote{\url{http://skyserver.sdss.org/dr12/en/tools/explore/summary.aspx?id=0x112d1f06c01f0a28}},
barely visible in the SDSS DR12 images.
Since it has no cataloged redshift, we measure its redshift via \species{O}{ii}
$\lambda 3727$ and \species{O}{iii} $\lambda\lambda 4960,5008$ lines in the
Keck I spectrum. The redshift is \num{0.0684 \pm 0.0001}, corresponding to a
luminosity distance of $\SI{\sim 300}{Mpc}$ assuming standard cosmology with
Hubble parameter $H_0 = \SI{70}{km.\second^{-1}.Mpc^{-1}}$, matter density
$\Omega_m = 0.3$, and dark energy density $\Omega_\Lambda = 0.7$.

Adopting the luminosity distance from above, the host galaxy has absolute
magnitudes of $M_g \approx \SI{-16.2}{mag}$, $M_r \approx \SI{-16.6}{mag}$ and
$M_i \approx \SI{-16.7}{mag}$\footnotemark[\value{footnote}] (corrected for
Galactic extinction).
Using the
luminosity-metallicity relation from \citet[eq.~1]{Lee2006}, we
find a metallicity of $12 + \log{\text{O}/\text{H}} \approx 8$,
indicating that the host galaxy is quite metal-poor.
Overluminous SNe IIn, such as PTF12csy, have been preferentially found to occur
in subluminous, low-metallicity galaxies \citep{Miller_2008iy,Stoll_2010jl},
such as the host of PTF12csy.
This is a trend, which is also observed for long GRBs
\citep{Stoll_2010jl}.
Statistics are still low and \citet{Miller_2008iy} cautioned that there could be
some selection bias due to intrinsically bright SNe in faint host galaxies being
more easily discovered during surveys doing aperture photometry. However, new
surveys performing image subtraction and observing large untargeted fields, e.g.
PTF and Pan-STARRS, provide increasing evidence for this trend, as most of the
discovered bright objects would have been luminous enough to be detected in bright
galaxies and in searches that are targeted to bright galaxies
\citep{Stoll_2010jl}.
PTF12csy, probably discovered by coincidence in an unbiased way, confirms this
emerging trend as well.

The SDSS DR12\footnotemark[\value{footnote}] has the host galaxy's
cataloged center position
about \SI{3}{\arcsec} away from the found SN position (see also
Fig.~\ref{fig:subtraction}).
With an apparent diameter of about \SI{5}{\arcsec}, corresponding to $\sim
\SI{7}{kpc}$, this is quite far from the center of the galaxy, i.e.\
about \SI{4}{kpc} off-center.

\section{SUMMARY AND CONCLUSION}
\label{sec:conclusion}

The highest significance alert from the IceCube Optical Follow-Up
program led to a coincidental discovery of the interesting and unusual
Type IIn SN PTF12csy, which was already at least 169 days old. The
combined \emph{a posteriori} significance of the neutrino doublet alert and the
coincident detection of any core-collapse SN within the error radius of the
neutrino events (\SI{0.54}{\degree}) and within the luminosity distance of the
SN (\SI{300}{Mpc}) is \SI{2.2}{\ensuremath{\sigma}}, for the time interval of
the IceCube data acquisition season 2011/12.

PTF12csy is rare and unusual: With peak absolute magnitudes of $M_r < -19$,
perhaps about -20, it belongs to the most luminous SNe. The SN is most likely
powered by interaction of the ejecta with a dense circumstellar medium (CSM).
The spectrum indicates a complicated structure of the CSM\@. Its host galaxy is a
faint and metal-poor dwarf galaxy, confirming an observed trend for luminous SNe
IIn. PTF12csy is similar in photometry and spectroscopy to other rare luminous
SNe IIn, e.g.\ SNe 2008iy and 2010jl. The total radiated energy is
\SI{2e50}{erg} within the first 400 rest frame days after detection.

Given the ejecta-CSM interaction, high-energy (HE) cosmic ray production and
neutrino emission may be expected on a time scale of \SIrange{1}{10}{months},
according to \citet{murase-IIn} and \citet{Katz2012}. However, the SN is too
far away for IceCube to detect this emission.  A complementary neutrino
analysis performed offline, using one year of IceCube data which cover
most of the optical SN fluence, does not reveal a signal-like accumulation of
neutrino events from the SN's position, leading to a very high upper limit of
more than 1000 times the tested model fluence, owing to the large distance.

Due to the long delay of several months between explosion date and neutrinos,
the doublet of neutrinos within less than two seconds cannot be explained by the
formation of a jet shortly after core-collapse according to the choked jet model
\citep{AndoBeacom2005}. Nor can it be explained by the expected HE neutrino
production from ejecta-CSM interaction of SNe IIn (s.a.). We therefore conclude
that the only reasonable explanation is that the SN detection was coincidental
and the neutrino doublet was produced by uncorrelated background events of
atmospheric neutrinos and/or mis-reconstructed atmospheric muons\footnote{with a
small statistical chance on the order of a few percent that one of the neutrinos
is part of the measured diffuse astrophysical flux, see \citet{IceCubeHESE}}.

However, if there was a delaying mechanism at work, such as the
spin-down of a supramassive neutron star that delays its collapse to a
black hole, as in the blitzar model \citep{Blitzar_Falcke}, then the
neutrino doublet may have originated from a jet forming in the course of
this delayed collapse few hundred days after SN explosion.

The coincidental detection of a Type IIn SN following an
IceCube neutrino alert demonstrates the capability of the follow-up
system to reveal transient HE neutrino sources. An advantage of the
follow-up paradigm is the prompt availability of multi-messenger
information for the identification of the source, as well as the mere
statistical significance of a coincidence between a neutrino burst and
an electromagnetic transient detection. In this case, the significance
is very low due to the delay of several months between explosion and
neutrinos. However, this coincidence motivates the continuation of the
follow-up program as well as further stacked neutrino analyses of Type
IIn supernovae.

\acknowledgments{

We acknowledge the support from the following agencies:
U.S. National Science Foundation-Office of Polar Programs,
U.S. National Science Foundation-Physics Division,
University of Wisconsin Alumni Research Foundation,
the Grid Laboratory Of Wisconsin (GLOW) grid infrastructure at the University of Wisconsin - Madison, the Open Science Grid (OSG) grid infrastructure;
U.S. Department of Energy, and National Energy Research Scientific Computing Center,
the Louisiana Optical Network Initiative (LONI) grid computing resources;
Natural Sciences and Engineering Research Council of Canada,
WestGrid and Compute/Calcul Canada;
Swedish Research Council,
Swedish Polar Research Secretariat,
Swedish National Infrastructure for Computing (SNIC),
and Knut and Alice Wallenberg Foundation, Sweden;
German Ministry for Education and Research (BMBF),
Deutsche Forschungsgemeinschaft (DFG),
Helmholtz Alliance for Astroparticle Physics (HAP),
Research Department of Plasmas with Complex Interactions (Bochum), Germany;
Fund for Scientific Research (FNRS-FWO),
FWO Odysseus programme,
Flanders Institute to encourage scientific and technological research in industry (IWT),
Belgian Federal Science Policy Office (Belspo);
University of Oxford, United Kingdom;
Marsden Fund, New Zealand;
Australian Research Council;
Japan Society for Promotion of Science (JSPS);
the Swiss National Science Foundation (SNSF), Switzerland;
National Research Foundation of Korea (NRF);
Danish National Research Foundation, Denmark (DNRF)

This paper is based on observations obtained with the
Samuel Oschin Telescope as part of the Palomar Transient Factory
project, a scientific collaboration between the
California Institute of Technology,
Columbia University,
Las Cumbres Observatory,
the Lawrence Berkeley National Laboratory,
the National Energy Research Scientific Computing Center,
the University of Oxford, and the Weizmann Institute of Science.
Some of the data presented herein were obtained at the W. M. Keck
Observatory, which is operated as a scientific partnership among the
California Institute of Technology, the University of California,
and NASA; the Observatory was made possible by the generous
financial support of the W. M. Keck Foundation.  We are grateful for
excellent staff assistance at Palomar, Lick, and Keck Observatories.
E.O.O.\ is incumbent of
the Arye Dissentshik career development chair and
is grateful to support by
grants from the
Willner Family Leadership Institute
Ilan Gluzman (Secaucus NJ),
Israeli Ministry of Science,
Israel Science Foundation,
Minerva and
the I-CORE Program of the Planning
and Budgeting Committee and The Israel Science Foundation.

PAE Acknowledges support from the UK Space Agency. This work made use of data
supplied by the UK Swift Science Data Centre at the University of Leicester.

The Pan-STARRS1 Surveys (PS1) have been made possible through contributions of
the Institute for Astronomy, the University of Hawaii, the Pan-STARRS Project
Office, the Max-Planck Society and its participating institutes, the Max Planck
Institute for Astronomy, Heidelberg and the Max Planck Institute for
Extraterrestrial Physics, Garching, The Johns Hopkins University, Durham
University, the University of Edinburgh, Queen's University Belfast, the
Harvard-Smithsonian Center for Astrophysics, the Las Cumbres Observatory Global
Telescope Network Incorporated, the National Central University of Taiwan, the
Space Telescope Science Institute, the National Aeronautics and Space
Administration under Grant No. NNX08AR22G issued through the Planetary Science
Division of the NASA Science Mission Directorate, the National Science
Foundation under Grant No. AST-1238877, the University of Maryland, and Eotvos
Lorand University (ELTE).

SJS acknowledges (FP7/2007-2013)/ERC grant agreement n$^{\rm o}$ [291222].

Funding for SDSS-III has been provided by the Alfred P.~Sloan Foundation, the
Participating Institutions, the National Science Foundation, and the U.S.
Department of Energy Office of Science. The SDSS-III web site is
\url{http://www.sdss3.org/}.

SDSS-III is managed by the Astrophysical Research Consortium for the
Participating Institutions of the SDSS-III Collaboration including the
University of Arizona, the Brazilian Participation Group, Brookhaven National
Laboratory, Carnegie Mellon University, University of Florida, the French
Participation Group, the German Participation Group, Harvard University, the
Instituto de Astrofisica de Canarias, the Michigan State/Notre Dame/JINA
Participation Group, Johns Hopkins University, Lawrence Berkeley National
Laboratory, Max Planck Institute for Astrophysics, Max Planck Institute for
Extraterrestrial Physics, New Mexico State University, New York University,
Ohio State University, Pennsylvania State University, University of Portsmouth,
Princeton University, the Spanish Participation Group, University of Tokyo,
University of Utah, Vanderbilt University, University of Virginia, University
of Washington, and Yale University.

We would like to thank K.~Murase and J.~Hjorth for helpful discussions.}






\end{document}